\documentclass[sigconf]{acmart}

\AtBeginDocument{%
  \providecommand\BibTeX{{%
    \normalfont B\kern-0.5em{\scshape i\kern-0.25em b}\kern-0.8em\TeX}}}

\setcopyright{none} 
\renewcommand\footnotetextcopyrightpermission[1]{} 

\usepackage{hyperref}
\usepackage{graphicx}
\usepackage[T1]{fontenc}
\usepackage{color}
\usepackage{booktabs}
\usepackage{amsmath}
\usepackage{multirow}
\usepackage{array}
 \usepackage{algorithm}
\usepackage{algorithmicx}
\usepackage[scientific-notation=true]{siunitx}
\usepackage{algpseudocode}
\usepackage{amsfonts}
\usepackage{subfig}
\usepackage{soul}

\newcolumntype{L}[1]{>{\raggedright\let\newline\\\arraybackslash\hspace{0pt}}m{#1}}
\newcolumntype{C}[1]{>{\centering\let\newline\\\arraybackslash\hspace{0pt}}m{#1}}
\newcolumntype{R}[1]{>{\raggedleft\let\newline\\\arraybackslash\hspace{0pt}}m{#1}}

\newcommand{\cmt}[2]{{\bf #1}: \textcolor{blue}{#2}}

\begin{document}

\title[ChASE - A Distributed Hybrid CPU-GPU Eigensolver for Hermitian Problems]{ChASE - A Distributed Hybrid CPU-GPU Eigensolver for Large-scale Hermitian Eigenvalue Problems}

\author{Xinzhe Wu}
\orcid{0000-0001-5716-3116}
\affiliation{%
  \institution{J\"{u}lich Supercomputing Centre~\\Forschungszentrum J\"{u}lich}
  \city{J\"{u}lich}
  \country{Germany}
  \postcode{52425}
}
\email{xin.wu@fz-juelich.de}

\author{Davor Davidovi\'c}
\orcid{0000-0003-2649-9236}
\affiliation{%
  \institution{Centre for Informatics and Computing~\\ Ru\dj er Bo\v skovi\' c Institute}
  \city{Zagreb}
  \country{Croatia}
  \postcode{10000}}
\email{ddavid@irb.hr}

\author{Sebastian Achilles}
\orcid{0000-0002-1943-6803}
\affiliation{%
  \institution{J\"{u}lich Supercomputing Centre~\\Forschungszentrum J\"{u}lich}
  \city{J\"{u}lich}
  \country{Germany}
  \postcode{52425}
}
\email{s.achilles@fz-juelich.de}

\author{Edoardo Di Napoli}
\orcid{0000-0001-5821-5897}
\affiliation{%
  \institution{J\"{u}lich Supercomputing Centre~\\Forschungszentrum J\"{u}lich}
  \city{J\"{u}lich}
  \country{Germany}
  \postcode{52425}
}
\email{e.di.napoli@fz-juelich.de}

\begin{abstract}
As modern massively parallel clusters are getting larger with beefier compute nodes, traditional parallel eigensolvers, such
as direct solvers, struggle keeping the pace with the hardware evolution and being able to scale efficiently 
due to additional layers of communication and synchronization.
This difficulty is especially important when porting traditional libraries to heterogeneous
computing architectures equipped with accelerators, such as Graphics Processing Unit (GPU). 
Recently, there have been significant scientific contributions to the development of 
filter-based subspace eigensolver to compute partial eigenspectrum. The simpler structure of these type of algorithms makes for them easier to avoid the communication and synchronization bottlenecks typical of direct solvers. The Chebyshev Accelerated Subspace Eigensolver (ChASE) is a modern subspace 
eigensolver to compute partial extremal eigenpairs of large-scale Hermitian eigenproblems
with the acceleration of a filter based on Chebyshev polynomials. In this work, we extend our previous work on  ChASE by adding support for distributed hybrid CPU-multi-GPU computing architectures.  
Out tests show that ChASE achieves very good scaling performance up to 144 nodes with 526 NVIDIA A100 GPUs in total on dense eigenproblems of size up to $360$k.
\end{abstract}

\ccsdesc[500]{Computing methodologies~Parallel algorithms}
\ccsdesc[500]{Mathematics of computing~Mathematical software performance}

\keywords{Subspace iteration eigensolver,
Dense Hermitian matrix,
Chebyshev polynomial,
Distributed hybrid CPU-GPU,
Heterogeneous GPU supercomputers}

\maketitle

\section{Introduction}

Modern scientific and engineering applications (e.g., in the filed of electronic structure in condensed matter physics  \cite{kohn1965self,becke1993new,kohn1999nobel}) often require
the solution of very large dense eigenproblems distributed on 
massive supercomputers.
In the last decades, there have been continuous 
efforts to develop efficient 
parallel eigensolver libraries for large dense matrices targeted at systems with a large number of compute nodes \cite{Choi2003ScaLAPACK:Computers, Poulson2013Elemental:Computations, ELPA2014EigenvalueELPA, imamura2011development}. Dense eigensolvers, even those for extremal eigenproblems in which only a fraction at the end of the spectrum is sought after, have $\mathcal{O}(n^3)$ complexity. 
Unlike simpler linear algebra operations, eigensolvers consist of several "moving parts", many of which can differ significantly in terms of computational intensity.
Because of these characteristics, it is often challenging to construct parallel implementations of dense eigensolvers that scale well on large supercomputers, 
and ensure a good balance between different nodes throughout the computation.
When it comes to efficiently porting dense eigensolvers to distributed GPGPU (General Purpose GPU) systems, this challenge becomes even harder to address. Currently, only the ELPA library~\cite{ELPA2014EigenvalueELPA,Yu2021GPU-accelerationEigenproblems} 
carried out such an endeavor with moderate success. 
To our knowledge, no library is able to successfully exploit GPUs for very large dense eigenproblems with size larger than 100k.

The lack of balanced scalability on heterogeneous platforms is an important issue in light of the current trend towards massively parallel clusters trying to reach exascale. To grasp this target, future 
architectures will have to leverage compute nodes equipped with
beefy multi-core CPUs coupled with powerful multi-GPUs via a high-bandwidth interconnect (e.g.
the NVIDIA Grace project \cite{gpu-grace}). Having 
a dense eigensolver running efficiently on such architectures is paramount. 
In order to develop efficient distributed multi-GPU eigensolver libraries, the implementations 
should be designed with a large amount of concurrency
and a minimal data transfer between host and device memory. 

To address this problem, we propose a distributed hybrid CPU-GPU version of the ChASE eigensolver.
ChASE, short for Chebyshev Accelerated Subspace iteration Eigensolver, 
is a modern library based on subspace iteration accelerated with a Chebyshev polynomial filter and includes some innovative algorithmic features~\cite{Berljafa:2014jv,Winkelmann2019}.
It is designed to approximate extremal eigenpairs of dense symmetric and Hermitian matrices and is particularly effective in solving sequences of correlated eigenproblems (e.g., derived by the linearization of non-linear
problems). 
An MPI-based parallel implementation of ChASE
for homogeneous systems has already been presented in \cite{Winkelmann2019}, 
and will be referred to as ChASE-CPU. 
We have referred to the hybrid CPU-GPU implementation presented in this paper as ChASE-GPU.

Due to its algorithm design, ChASE is able to scale well on distributed multi-GPU clusters. 
As shown in \cite{Winkelmann2019}, the algorithm can be decoupled in a series of basic linear operations, 
the most important of which is the
Hermitian Matrix-Matrix Multiplications ({\tt HEMM}s) repeatedly executed within the Chebyshev polynomial Filter.
As a typical {\tt BLAS-3} operation, the performance of an efficient implementation of the distributed {\tt HEMM} 
is able to approach the theoretical peak performance of a given system.
Other linear algebra operations are computed redundantly on each MPI process, minimizing the inter-node communication.
We provide the hybrid CPU-GPU implementation of ChASE by designing and
implementing a customized distributed {\tt HEMM} that supports flexible configurations
of binding MPI processes and GPUs within the compute node.
In addition, selected computationally intensive linear algebra operations, such as
QR factorization, are also offloaded to one of the GPUs tied to each MPI process.
Because to the relatively small memory capacity of the device, we provide accurate formulas for estimating the memory cost for CPU and GPU to help the user choose the optimal resource usage for a given problem.
ChASE-GPU has been
tested on our in-house supercomputer \textsc{JURECA Data Centric Module} (JURECA-DC).
The strong scaling tests were performed with a symmetric matrix of size 130k with double-precision using up to 64 compute nodes and with a total of 256 NVIDIA A100 GPUs. Similarly, the weak scaling test was performed using up to 144 compute nodes with a total of 576 GPUs, using symmetric matrices ranging in size from 30k to 360k. We have also performed a strong scaling test up to 64 compute nodes comparing ChASE-GPU with ELPA2 with GPU support. This last test was carried out on an Hermitian eigenproblem of
size 76k generated by the discretization of the Bethe-Salpeter equation used to simulate the opto-electronic properties of In$_2$O$_3$.

\textit{Original contributions.}\ 
We have significantly improved the performance of the existing custom-HEMM, which is mainly tailored to the execution of the 3-terms recurrence relation at the base of the polynomial Filter. Although ChASE-CPU already supported GPUs, it could only use a single GPU per MPI rank. We increased the flexibility of the custom-HEMM by extending support for multiple GPUs per MPI rank. We achieved this result by adding a further custom distribution of the data to allow the execution of multi-GPU HEMM operations local to each MPI rank. We also optimized the design of the filter by removing redundancies in the code and reducing the memory footprint of the algorithm.  
As a result we gained an increased scalability of the polynomial Filter in terms of the number of CPUs and GPUs, 
and ensured that much larger eigenproblems can be solved efficiently with the same amount of resources by making well-balanced use of all available computational units.

\textit{Organization.}\ In Section \ref{Distributed eigensolvers}, we give a short overview of existing dense symmetric and Hermitian eigensolvers targeting distributed memory architectures followed by a description of
the ChASE algorithm and its detailed implementation on distributed multi-GPUs, along with formulas for the memory requirements of CPU and GPU.
We present the numerical and parallel performance of ChASE-GPU in Section \ref{results} and a comparison with currently available eigensolvers executing on distributed GPUs. Section \ref{Conclusion} summarizes the achievements and concludes the paper.

\section{Distributed eigensolvers}\label{Distributed eigensolvers}

In this paper, we look for a partial diagonalization of {\sf nev} eigenpairs of a standard symmetric eigenvalue problem

\begin{equation}
    AV=V\Lambda,
\end{equation}

in which $A$ is a $n\times n$ real symmetric or complex Hermitian matrix. $V$ is a $n\times {\sf nev}$ rectangular matrix, and $\Lambda$ is a ${\sf nev}\times {\sf nev}$ diagonal matrix, which contain the desired {\sf nev} eigenvectors and eigenvalues, respectively.
While the matrix $A$ can be sparse, dense or banded, 
this paper focuses only on dense matrices. Depending on the number of eigenpairs
to be computed, eigenproblems can be solved either by \textit{direct solvers} or \textit{iterative solvers}. Direct solvers are generally used when many if not all eigenpairs are needed. Iterative methods are more effective when {\sf nev} is a small fraction of $n$. Direct solver generally consists of three phases: (1) reducing original matrix to a condensed form (usually tridiagonal form{, but other banded forms~\cite{imamura2011development,fukaya2015performance} also exist}) by orthogonal transformation; (2) solving eigendecomposition of this condensed form through QR {algorithm~\cite{sameh1977parallel}},
divide-and-conquer method~{\cite{tisseur1999parallel}, MRRR algorithm~\cite{dhillon2006design}}, etc;  and (3) backtransforming to obtain the eigenvectors of the original matrix, if required. On the other hand, iterative methods project the eigenproblem onto a small continuously improved searching space. Then, an approximated basis for desired eigenvectors can be constructed within this small searching space
to extract desired eigenpairs. Convergence of iterative methods highly depends
on the spectrum of $A$, therefore filtering (e.g., polynomial or rational filters) and preconditioning techniques have been proposed {\cite{ashby1991minimax,sakurai2003projection,thornquist2006fixed,polizzi2009density,ikegami2010filter}}, which are able to separate the desired eigenpairs in a specific range from the rest.

Numerous libraries providing the distributed-memory eigensolvers, especially the direct solvers, have been available for
the last decades, since the first release of the ScaLAPACK~\cite{Choi2003ScaLAPACK:Computers} library. ScaLAPACK extends the ubiquitous LAPACK~\cite{Anderson1999LAPACKGuide} re-implementing its routines by dividing the matrices into blocks and distributing them into 2D block-cyclic fashion. Although ScaLAPACK brings a good level of scalability and performance on distributed CPU-only systems, it cannot exploit modern heterogeneous systems based on accelerators. In recent years, additional cutting-edge direct solver libraries have been introduced, such 
as ELPA~\cite{marek2014elpa,ELPA2014EigenvalueELPA} and EigenEXA~\cite{Fukaya2018AComputer}. These new libraries employ a similar 2D block-cyclic scheme, with further optimizations for node-level and distributed-memory level operations and communications, and achieve better performance than ScaLAPACK. Despite being out of maintenance since 2016, the Elemental~\cite{Poulson2013Elemental:Computations} library 
implements distributed direct solvers with a cyclic-cyclic data distribution and a parallel direct solver based on the MRRR algorithm {\cite{bientinesi2005parallel}}. 
Among the iterative libraries, the most well-known solvers for dense symmetric problem are FEAST~\cite{polizzi2009density}, and its distributed-memory variant PFEAST~\cite{kestyn2016pfeast}. As a typical subspace method, 
PFEAST projects eigenproblems onto a set of subspaces constructed by rational filters.

To our knowledge, though numerous solvers for dense eigenproblems have been developed for distributed-memory systems, only a few of them can exploit distributed hybrid systems equipped with GPUs. Recently, significant work has been conducted on extending distributed-memory eigensolvers to support GPU acceleration~\cite{Haidar2013,Myllykoski2020}. The most recent version of   ELPA2~\cite{Yu2021GPU-accelerationEigenproblems} introduces eigensolvers capable of exploiting distributed heterogeneous systems equipped with GPUs.
Considered to be the future replacement
for ScaLAPACK, SLATE~\cite{Gates2019SLATE:Library} is a cutting-edge library providing dense
linear algebra routines supporting large-scale distributed-nodes with accelerators. Currently, SLATE supports only
the computation of eigenvalues and lacks the backtransformation functionality required to compute the eigenvectors of the original matrix.
In \cite{Williams-Young2020}, authors implemented a Shift-Invert Spectrum Slicing subspace eigensolver based on GPU-accelerated dense linear algebra kernels in SLATE.
There are other GPU-specialized libraries, such as cuSOLVER (single GPU)~\cite{NVIDIA2019}, cuSOLVER-MG (multi-GPUs)~\cite{NVIDIA2019}, and MAGMA (hybrid CPUs and multi-GPUs)~\cite{Magma}, which provide GPU-accelerated direct solvers. However, they are all tailored for shared-memory systems only, and therefore the eigenproblem size they can tackle is limited by the size of the device memory on the compute node.

With this paper we present a distributed hybrid CPU-GPU implementation of ChASE and propose it as
an alternative for solving large symmetric (Hermitian) eigenproblems that go beyond the state of the art. With the acceleration of the Chebyshev polynomial filter, ChASE makes it extremely efficient to approximate partial extremal eigenpairs ($<10$\%). ChASE is highly scalable, because most of its work is
concentrated in Matrix-Matrix multiply. With these {\tt BLAS-3 HEMM} kernels, ChASE capitalizes
on their extreme effectiveness to achieve a high efficiency both on each GPU card of compute node and in a distributed-memory architecture.
\section{Distributed multi-GPU ChASE}\label{ChASE}

\subsection{ChASE Algorithm}

ChASE is a numerical library written in C++ with templates and based on the subspace iteration algorithm. Subspace iteration is one of the first iterative algorithms used as numerical eigensolver for Symmetric/Hermitian matrices~\cite{Bauer:1957hv}. Its more sophisticate cousin, complemented with a Chebyshev polynomial filter, was developed quite early on as a numerical code by Rutishauser ~\cite{Rutishauser:1970wc}. In early 2000s, a version was developed to solve electronic structure eigenproblems within the PARSEC code
~\cite{Zhou:2006ut,Zhou:2014fe}.  

Last ten years have seen a revival of this algorithm in the context of applications mostly focused on realizations of electronic structure codes~\cite{Berljafa:2014jv,Levitt:2015wc,Banerjee:2016fy}. The ChASE library evolved from one of these efforts and became a full fledged numerical eigensolver which can be used outside the specific 
electronic structure domain~\cite{Winkelmann2019}. ChASE's algorithm takes inspiration by the work of Rutishauser~\cite{Rutishauser:1970wc} and Zhou et al.~\cite{Zhou:2006ut}, and includes some additional features: 1) it introduces an internal loop that iterates over the polynomial filter and the Rayleigh quotient, 
2) it uses a sophisticated mechanism to estimate the spectral bounds of the search subspace using a Density of States (DoS) method~\cite{Lin:2013to}, 3) it adds a deflation and locking mechanism to the internal loop, and 4) most importantly, it optimizes the degree of the polynomial filter so as to minimize the number of matrix-vector operations required to reach convergence of desired eigenpairs. 

Full details of ChASE structure and its algorithm can be found in~\cite{Winkelmann2019}, here we give a high level description of its main parts (see Algorithm~\ref{alg:chase})
ChASE first estimates the necessary spectral bounds by executing a small number of repeated {\tt Lanczos} steps (Line \ref{alg:chase:lanczos}). 
It then filters a number of (random) vectors using an optimized degree for each vector (Line \ref{alg:chase:filter}). The filtered vectors are orthonormalized using QR factorization (Line \ref{alg:chase:qr}). 
The $Q$ factor 
is used to reduce the eigenproblem to the size of the subspace yet to be diagonalized using a {\tt Rayleigh-Ritz} projection (Line \ref{alg:chase:rr}). The resulting ``small'' eigenproblem is solved using a standard dense solver (e.g. Divide\&Conquer). Residuals are then computed and eigenpairs below the tolerance threshold are deflated and locked (Line \ref{alg:chase:resid}). Finally, a new set of filtering degrees are computed for the non-converged vectors and the procedure is repeated (Line \ref{alg:chase:degrees}). 

\begin{algorithm}[t]
  \caption{ChASE algorithm}\label{alg:chase}
  \begin{algorithmic}[1]
    \Require {Symmetric matrix $A$, number of desired eigenpairs {\sf nev}, threshold tolerance for residuals $tol$, initial polynomial degree $deg$, search space increment $nex$, vector $\hat{V}\equiv[\hat{v}_1,\cdots,\hat{v}_{{\sf nev}+{\sf nex}}]$.}
    \Ensure {{\sf nev} extreme eigenpairs $(\Lambda, \hat{Y})$, with $\Lambda=[\lambda_1,\cdots,\lambda_{{\sf nev}}]$ and $\hat{Y}\equiv[\hat{y}_1,\cdots,\hat{y}_{{\sf nev}}]$.}
    \State $m_{1:{\sf nev}+{\sf nex}}\leftarrow deg$
    \State $(b_{sup},\mu_1,\mu_{{\sf nev}+{\sf nex}},\hat{V})\leftarrow$ \texttt{Lanczos}$(A)$ \label{alg:chase:lanczos}
    
    \While {size($\hat{Y}<${\sf nev})}\label{alg:chase:while}
        \State $\hat{V}\leftarrow$ \texttt{Filter}($A, b_{sup}, \mu_1, \mu_{{\sf nev}+{\sf nex}},\hat{V}, m$)\label{alg:chase:filter}
        \State $\hat{Q}\leftarrow$ \texttt{QR}($[\hat{Y} \, \hat{V}]$)\label{alg:chase:qr}
        \State $(\hat{V},\tilde{\Lambda})\leftarrow$ \texttt{Rayleigh-Ritz}$(A,\hat{Q})$\label{alg:chase:rr}
        \State Compute the \texttt{residual} $Res(\hat{V},\tilde{\Lambda})$\label{alg:chase:resid}
        \State $(\hat{V}, \Lambda, \hat{Y})\leftarrow$ \texttt{Deflation \& Locking}($\hat{V},$$\tilde{\Lambda},$ $Res(\hat{V},\tilde{\Lambda}),$$\hat{Y}$)\label{alg:chase:deflated}
        \State $\mu_1\leftarrow\min([\Lambda \, \hat{\Lambda}])$, $\mu_{{\sf nev}+{\sf nex}}\leftarrow\max([\Lambda \, \hat{\Lambda}])$
        \State $c\leftarrow\frac{b_{sup}+\mu_{{\sf nev}+{\sf nex}}}{2}$, $e\leftarrow\frac{b_{sup}-\mu_{{\sf nev}+{\sf nex}}}{2}$
        \For{$a\in [1,\cdots, size(\hat{V})]$}
            \State $m_a\leftarrow \texttt{Degrees}(tol, Res(\hat{V}_{:,a},\tilde{\lambda}_a),\lambda_a,c,e)$\label{alg:chase:degrees}
        \EndFor
        \State \texttt{Sort} $Res(\hat{V},\tilde{\Lambda})$, $\hat{V}$, $\tilde{\Lambda}$, $m$ according to $m$
    \EndWhile\label{alg:chase:endwhile}
  \end{algorithmic}
\end{algorithm}

\subsection{Distributed Implementation of ChASE}\label{subsec:distChase}

The implementation of ChASE relies on a number of numerical kernels which can be further 
decoupled as simple dense linear algebra operations to exploit optimized BLAS and LAPACK
libraries (e.g., MKL {\cite{wang2014intel}}, OpenBLAS {\cite{xianyi2012openblas}}, BLIS {\cite{van2015blis}}, libFLAME {\cite{van2009libflame}}).
Such decoupling makes easy for ChASE to take
advantage of low-level kernels. In \cite{Winkelmann2019}, the authors introduce a distributed implementation of ChASE in which most operations such as QR factorization and the eigendecomposition within the \texttt{Rayleigh-Ritz} section have been implemented with vendor-optimized threaded BLAS/LAPACK. The only exception is the Hermitian
matrix-matrix multiplication ({\tt HEMM}), which occupies a significant part
of computations within ChASE. This {\tt HEMM} is implemented with a custom MPI scheme, and is used in the \texttt{Filter}, \texttt{Rayleigh-Ritz}, and \texttt{Residual} parts of the ChASE Algorithm whenever the matrix $A$ is right-multiplied
by a rectangular matrix $\hat{V}$.

In our customized distributed {\tt HEMM}, MPI processes are organized in a 2D
grid whose shape is as square as possible. Matrix $A$ is divided into
sub-blocks, each of which is assigned onto one MPI process following the 2D block distribution. Within each row communicator, the rectangular matrix $\hat{V}$ is distributed in a 1D block fashion. This distribution results in a large and contiguous matrix-matrix multiplication on each node, often resulting in a performance close to its theoretical peak. The equation below gives an example which assigns a $n \times n$
matrix $A$ onto a $3 \times 2$ MPI grid. MPI processes are numbered using column-major order.

\begin{equation}
\label{eq:distribution:AV}
\begin{split}
    A_{dist}=\left(\begin{array}{c|c}
     A_{0,0} & A_{0,1}  \\
     \hline
     A_{1,0} & A_{1,1}  \\
     \hline
     A_{2,0} & A_{2,1}
   \end{array}\right)
\end{split},
\:
\begin{split}
    \hat{V}_{dist}=
    \left(\begin{array}{c|c}
     \hat{V}_{0} & \hat{V}_{1}  \\
     \hline
     \hat{V}_{0} & \hat{V}_{1}  \\
     \hline
     \hat{V}_{0} & \hat{V}_{1}
   \end{array}\right)
\end{split}
\end{equation}

In this example,
matrix $A$ is split in a series of sub-matrices $A_{i,j}$ with
$i \in [0,2]$ and $j \in [0,1]$. $A_{0,0}$ is assigned
to MPI rank 0, $A_{1,0}$ is distributed
to rank 1, and so on.
The rectangular matrix $\hat{V}$ is split horizontally into 2 sub-blocks $\hat{V}_0$ and $\hat{V}_1$, 
which satisfy the relation $\hat{V}^T=[\hat{V}_0^T \:|\: \hat{V}_1^T]$.
The distribution of $\hat{V}$ within each row communicator is also 
shown in Equation \ref{eq:distribution:AV}, in which $\hat{V}_0$ is assigned 
to the first column communicator---the MPI processes numbering $0$, $1$, and $2$---and $\hat{V}_1$ is assigned to the second column communicator.  

In the Chebyshev {\tt Filter} the matrix-matrix multiplications appears as a three-terms recurrence relation:

\begin{equation}
    \label{eq:three-terms-recurrence}
    \hat{V}_{i+1} = \alpha_i(A-\gamma_i I_n)\hat{V}_{i} + \beta_i \hat{V}_{i-1}, \quad i \in [1,m),
\end{equation}
where $\hat{V}$ is a (subset of) rectangular matrix, $m$ is the degree of Chebyshev polynomial, and $\alpha_i$, $\beta_i$, $\gamma_i$ are scalar parameters related to each iteration. As it is described, the {\tt HEMM} requires to re-distribute $\hat{V}$ from the iteration $i$ to $i+1$. Therefore, the matrix-matrix multiplication can be rewritten into two separate forms for iterations $i$ and $i+1$ as follows:
\begin{subequations}
\begin{align}
        \hat{W}=\hat{A}\hat{V},         \label{eq:w=av} \\
        \hat{V}=\hat{A}\hat{W}.         \label{eq:v=aw}
\end{align}
\end{subequations}
Here $\hat{A}=A-\gamma I_n$ and $\hat{V}$ are of
same distribution schemes as in Equation \ref{eq:distribution:AV}, 
and $\hat{W}=\hat{A}\hat{V}$ is a rectangular matrix with
the same size and shape of $\hat{V}$, but a 1D distribution scheme along the column communicator of the 2D grid. An example of $\hat{W}$ analogous to Equation \ref{eq:distribution:AV} is given below:

\begin{equation}
\label{eq:distribution:W}
\begin{split}
    \hat{W}_{dist}=\left(\begin{array}{c|c}
     \hat{W}_{0} & \hat{W}_{0}  \\
     \hline
     \hat{W}_{1} & \hat{W}_{1}  \\
     \hline
     \hat{W}_{2} & \hat{W}_{2}
   \end{array}\right),
\end{split}
\end{equation}
The $n\times n_e$ rectangular matrix $\hat{W}$ is split horizontally into 3
sub-blocks $\hat{W}_0$, $\hat{W}_1$ and $\hat{W}_2$, which
satisfy the relation $\hat{W}^T=[\hat{W}_0^T \: | \: \hat{W}_1^T \: | \: \hat{W}_2^T]$. For each iterative step in Equation \ref{eq:three-terms-recurrence}, it is necessary to either re-distribute $\hat{V}$ to $\hat{W}$
or \textit{vice versa}. As such, the communication cost of these frequent re-distributions
would clearly hamper the parallel performance of ChASE.

The scheme introduced in \cite{Winkelmann2019} avoids this re-distribution between Equation \ref{eq:w=av} and \ref{eq:v=aw} by right-multiplying the rectangular matrix on $\hat{A}^T$, the transpose of $\hat{A}$. This is possible since 
$\hat{A}$ is symmetric/Hermitian. Rectangular matrices are re-assembled on each MPI node via a broadcast operation within each column or row communicator after a full execution of a Chebyshev \texttt{Filter}, since other operations on these rectangular matrices are
performed redundantly on each MPI node. For more details of this scheme, please refer
to the paper \cite{Winkelmann2019}.   

\subsection{Multi-GPU Acceleration of ChASE} \label{Multi-GPU Acceleration of ChASE}

The ChASE-CPU algorithm~\cite{Winkelmann2019} allows easy offloading of the computational kernels to CPU or GPU using architecture-specific libraries. Currently, ChASE supports a single GPU per MPI rank per compute node, and only general matrix-matrix multiplication is offloaded to the GPU through a standard call to the CUDA HEMM kernel.
Since ChASE cannot exploit the full potential of distributed multi-GPU platforms, we have extended the current implementation with a customised multi-GPU {\tt HEMM}.

\subsubsection{Distributed Multi-GPUs HEMM}

\begin{figure}[t]
\centering
\subfloat[$\hat{W} = A * \hat{V} + \hat{W}$.]{\includegraphics[width=.95\linewidth]{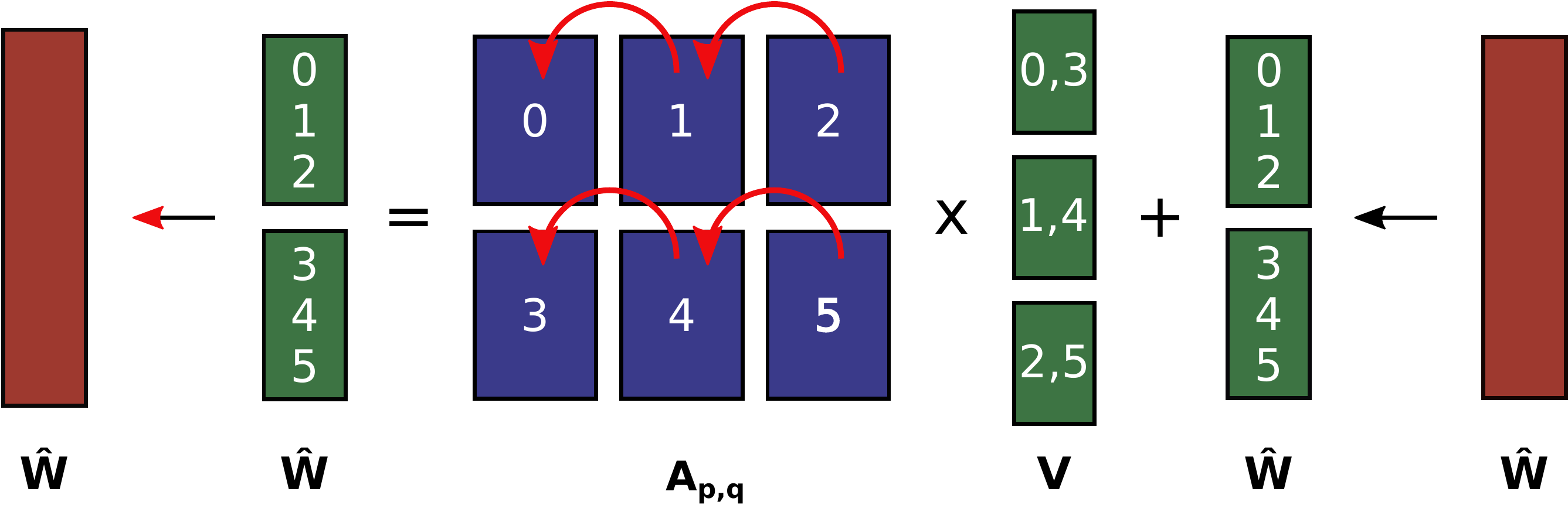}\label{fig:gpuHemmCAB}}
~\\
\subfloat[$\hat{V} = A^T * \hat{W} + \hat{V}$.]{\includegraphics[width=.95\linewidth]{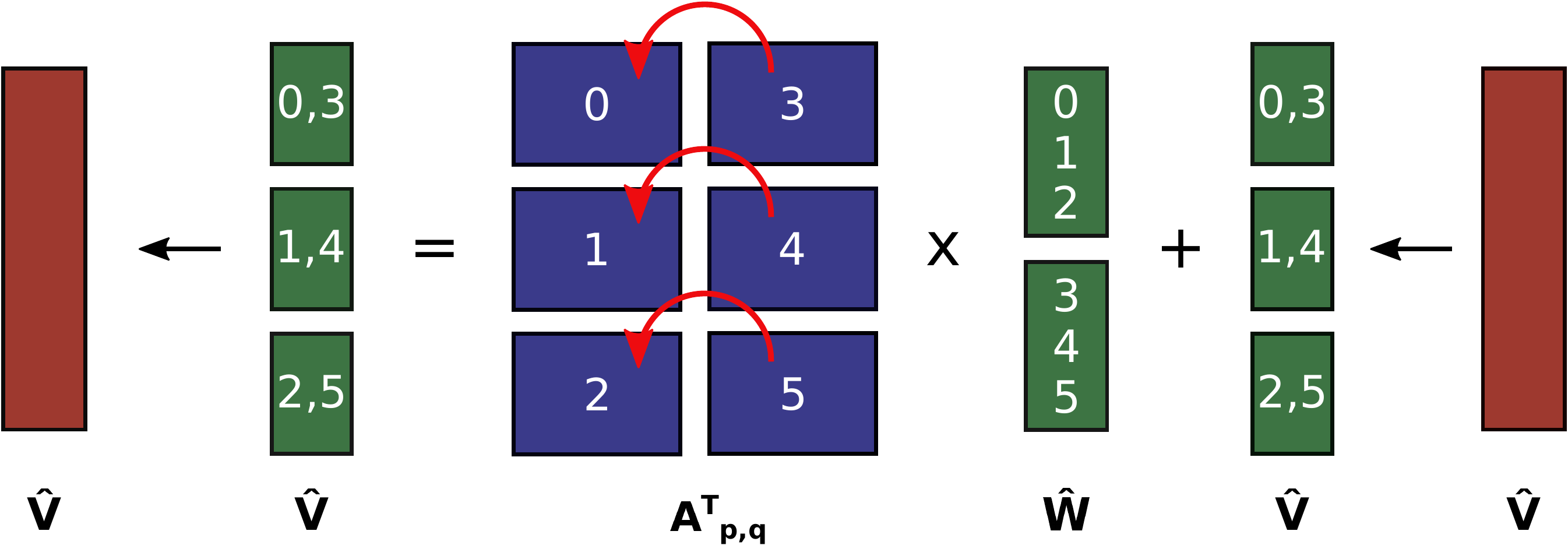}\label{fig:gpuHemmBAC}}

\caption{An example of the Multi-GPU {\tt HEMM} on 6 GPUs per MPI rank. 
Matrices/blocks stored in CPU (red) and GPU (green and blue) memory. The numbers denote the GPUs in which a block is stored.} \label{fig:multiGPUHEMM}
\end{figure}

Parallelizing {\tt HEMM} across multiple GPUs is done at two levels: 1) across MPI ranks and 2) across the available GPUs per MPI rank. The former is implemented in the same way as in ChASE- CPU (subsection~\ref{subsec:distChase}) by dividing the matrix A into blocks $A_{p,q}$ and distributing them over a 2D MPI grid (Eq.~\ref{eq:distribution:AV}, left).
At the MPI rank level, a dedicated block $A_{p,q}$ is further divided into several sub-blocks corresponding to the number of GPUs organized in the 2D grid. Fig.~\ref{fig:gpuHemmCAB} shows the subdivision of block $A_{p,q}$ (blue) and the rectangular matrices $\hat{V}, \hat{W}$ (green) on the GPU devices. The sub-blocks of $A_{p,q}$ are transmitted to the local GPUs only once and remain in GPU memory until ChASE completes. Since the entire input matrix $A$ is kept in the GPUs, the space required to store $A$ should fit in the aggregate memory of all available GPUs (see subsection~\ref{subsec:MemReq}). The rectangular matrices $\hat{V}$ and $\hat{W}$ are split as in Eq.~\ref{eq:distribution:AV} (right) and Eq.~\ref{eq:distribution:W}, respectively, and distributed among the GPUs according to the type of operation performed, as shown in Fig.~\ref{fig:multiGPUHEMM}.
The matrix-matrix product is executed on the GPUs in a block fashion. The communication between the GPUs is only at the node level (not via MPI) and along the rows of the 2D GPU grid. Upon completion, the resulting rectangular matrix $\hat{W}$ (Fig.\ref{fig:gpuHemmCAB}) is located in the first column communicator (e.g.~GPUs $0$ and $3$ in Fig.\ref{fig:gpuHemmCAB}).
The final step is to redistribute the obtained blocks of $\hat{W}$ across each row communicator, since the first block of $W$ needs to be distributed across GPUs $0$, $1$, and $2$ for the next iteration (see Fig.~\ref{fig:gpuHemmBAC}).
To reduce the unnecessary memory transfers and memory redistribution per MPI rank and between iterations of the \texttt{Filter}, the right-multiply is performed with $A_{p,q}^T$, Fig.~\ref{fig:gpuHemmBAC}.

Before performing the {\tt HEMM}, the matrix $A$ is shifted as $\hat{A}=A-\gamma I_n$ from the iteration $i$ to $i+1$ (see Section \ref{subsec:distChase}). We implemented specific CUDA kernels to efficiently carry out a new $\gamma$ shift of the matrix on each sub-blocks of $A$ for each GPU.  
This ensures these sub-blocks
reside always on the device memory of the GPUs without any data movement during the whole life-cycle of ChASE-GPU.

\subsubsection{Offloading Selected Routines to GPUs}

In ChASE-CPU, all dense linear algebra
operations other than {\tt HEMM} have been implemented
redundantly on each MPI process using threaded BLAS 
and LAPACK. For ChASE-GPU, the most computationally intensive operations 
have been offloaded to GPUs using NVIDIA 
shared-memory libraries cuBLAS and cuSOLVER. The API of cuBLAS and cuSOLVER
is almost identical to that of BLAS and LAPACK, hence offloading them
is a straightforward process. 

Because the communication between CPUs and GPUs should be minimized and the device memory of a single GPU is limited, only the most compute-intensive operations have been offloaded
to GPUs.
First we offload the QR factorization using
the cuSOLVER routine {\tt cusolverDnXgeqrf}. Other selected operations
are in the {\tt Rayleigh-Ritz} procedure.
In ChASE, the original eigenproblem 
is projected onto a "search" subspace, 
from which approximate solutions are computed. 
The active subspace is obtained by forming a ${\sf nev} \times {\sf nev}$ {\tt Rayleigh-Ritz} quotient
$G=\hat{Q}^TA\hat{Q}$, where $\hat{Q}$ is the $n\times {\sf nev}$ orthonormal matrix outputted by the QR factorization.
In ChASE, the right-multiplication of $\hat{Q}$ with $A$ is implemented by utilizing the available distributed multi-GPUs {\tt HEMM}.
The left-multiplying $\hat{Q}^T$ on $A\hat{Q}$ is 
offloaded to GPUs using the cuBLAS {\tt cublasXgemm} routine. 
Then $G$ is diagonalized as $G=\hat{W}\hat{\Lambda}\hat{W}^{-1}$ using a LAPACK eigensolver such as Divide\&Conquer, 
with ($\hat{W}, \hat{\Lambda}$) being the approximate eigenpairs of $G$. 
The diagonalization of $G$ is not performed on GPUs even if it would probably end up in a faster performance thanks to
less data movement between CPU and GPU. This design choice is deliberate and is based on the trade-off between the large memory requirement of the diagonalization and its relative small speedup over vendor-optimized LAPACK. 
The eigenvectors of the original problem are obtained
by a backtransform operation $\hat{Q}\hat{W}$, which is also offloaded to GPUs using
the {\tt cublasXgemm} routine.

Calling a cuBLAS or cuSOLVER routine requires allocating device memory for the 
input/output arrays and external workspace. This allocation 
is performed before the main work of ChASE by pre-computing the
required buffer size, which is then reused whenever is possible, and deallocated only after the main work ends. This implementation avoids frequent allocation and deallocation of device memory in the loop
between Line \ref{alg:chase:while} and \ref{alg:chase:endwhile} (Algorithm~\ref{alg:chase}),
which is important to reduce the CPU-GPU synchronizations. The data movement between 
host and device memory is limited since it takes place only once for each iteration within 
the main loop of ChASE. 
In future work we plan to explore GPU-aware MPI for the
direct communications between GPUs.

\subsection{Estimating Memory Requirement} \label{subsec:MemReq}

An important aspect of running ChASE is its memory footprint. 
The memory cost per task and per GPU device should not exceed the amount of 
main and device memory available. For this reason, we provide explicit formulas 
for estimating 
the memory cost of CPUs and GPUs in ChASE-GPU. The same formulas are encoded in a Python script (provided with the code) that the user can run to determine an appropriate resource allocation for a given problem.



The main memory requirement per MPI rank is given as follows

\begin{equation}
\label{eq:CPU_mem}
\begin{aligned}
M_{cpu}=pq+(p+q)n_e+2n_en,
\end{aligned}
\end{equation}
where $n$ is the rank of the matrix $A$
defining the eigenproblem, 
$n_e=\textsf{nev}+\textsf{nex}$ is the largest dimension of the active subspace to be projected onto. 
The dimension of the 2D MPI grid is defined as $r\times c$, {and the dimension of the local matrix held by
each MPI rank is $p\times q$, where $p=\frac{n}{r}$ and $q=\frac{n}{c}$.} Because of their dependence on $r$ and $c$, the first two terms of Equation~\ref{eq:CPU_mem} scale with the increase in computational resources, while the last term does not, since it refers to a part of the code that is redundantly executed.
The non-scalable part is negligible if
$n_e$ is a small percentage of $n$.


A similar expression holds for the memory requirement per GPU

\begin{equation}
\label{eq:GPU_mem}
\begin{aligned}
M_{gpu}=\frac{pq}{r_gc_g}+3\max(\frac{p}{r_g},\frac{q}{c_g})n_e +(2n+n_e)n_e,
\end{aligned}
\end{equation}
In ChASE-GPU, multiple GPUs of a single compute node can bind to an MPI process as a $r_g\times c_g$ 2D grid scheme.
The first two terms of Equation~\ref{eq:GPU_mem} also scale with resource allocation.
As for the CPU formula, the last term, which is $\mathcal{O}(n_en)$, mainly refers to the memory requirements of {{\tt cublasXgemm} 
and {\tt cusolverDnXgeqrf},
which are offloaded to GPUs and do not scale with the increase in MPI tasks}.
Since the capacity of device memory is limited compared to the main memory, this last term sets a maximum
size for matrix $A$ and the number of eigenpairs that can be computed.
In future work, we plan to remove these constraints by implementing versions of the 
related dense linear algebra operations distributed on a local subset of computing nodes.

When comparing ChASE memory footprint with the typical symmetric eigensolver in ScaLAPACK (e.g. {\tt PDSYEVX} based on parallel bisection and inverse iteration), we notice a similar leading order behavior.
{\tt PDSYEVX} requires $\mathcal{O}(\frac{n^2}{r c} \equiv pq)$ memory per processor (i.e. MPI rank)~\cite{Antonelli2005}. However, depending on the spectrum (e.g.~tightly clustered eigenvalues), the algorithm solving the tridiagonal form may require $\mathcal{O}(n^2)$ memory per processor to guarantee the correctness of the computed eigenpairs. So we conclude that, while in the general the ChASE CPU memory requirement is slightly larger than ScaLAPACK solvers and has a non-scalable portion that depends on the global dimension of A ($n$), this portion is not leading order. Moreover, because this portion is related to the redundant QR factorization, future development in distributing such factorization will significantly decrease its impact to the overall ChASE memory footprint.

\section{Numerical experiments}\label{results}

ChASE has been tested on JURECA-DC supercomputer at
J\"ulich Supercomputing Centre. Each node is equipped with 
two 64 cores AMD EPYC 7742 CPUs @ 2.25 GHz ($16 \times 32$ GB DDR4 Memory) and four NVIDIA Tesla A100 GPUs ($4 \times 40$ GB high-bandwidth memory). ChASE is compiled with GCC 9.3.0, OpenMPI 4.1.0 (UCX 1.9.0), CUDA 11.0 and Intel MKL 2020.4.304. All computations in this section are performed in double-precision.

\subsection{Test matrix suite}
\label{Random Matrices with Given Spectra}

For benchmarking ChASE, we use 
artificial matrices 
whose eigenpairs are known analytically and random matrices generated with given spectral properties. The framework for the matrix generation is inspired by the testing infrastructure for symmetric tridiagonal eigensolvers of LAPACK~\cite{marques2008algorithm}. In this work, double precision artificial matrices are generated with four different spectral distributions.


The first two generated matrices have analytical eigenvalues and are named \textsc{(1-2-1)} and \textsc{Wilkinson}. 
\textsc{(1-2-1)}
is a tridiagonal matrix with entries on the main diagonal and first two subdiagonals equal to $2$ and $1$, respectively.  
The \textsc{Wilkinson} matrix is
another tridiagonal matrix whose entries on the first subdiagonals are all
$1$, while the main diagonal have values $(m, m-1, m-2, \cdots, 2, 1, 2, \cdots, m-2, m-1, m)$, in which $m=\frac{n-1}{2}$ with $n$ the
size of matrix.

The next two generated matrices, \textsc{Uniform} and \textsc{Geometric}, are dense, symmetric with a given spectral distribution.
In order to generate them
we construct a diagonal matrix $D$ whose diagonal
is filled exactly by the prescribed eigenvalues. Then a dense matrix $A$ with the given spectra is generated as $A=Q^TDQ$, with $Q$ an orthogonal matrix, and $Q^T$ its transpose. The orthogonal matrix $Q$ is 
the Q factor of a QR factorization on a $n \times n$ matrix
whose entries are randomly generated with respect to the Gaussian distribution. 
All the matrices used in our tests are generated using our matrix  generator\footnote{https://github.com/SimLabQuantumMaterials/DEMAGIS} which allow the creation of matrices of any desired size for both shared-memory and distributed-memory architectures.


\begin{table}[t]
	\renewcommand{\arraystretch}{1.1}
	\caption{Spectral information for generating test matrices. In this table, we have $k = 1, \cdots,n$.}\label{spectral properties of generated matrices}
	\centering
	\begin{tabular}{C{3.1cm}|C{4.5cm}}
		\toprule
		Matrix Name & Spectral Distribution  \\
		\midrule
		\textsc{Uniform} (\textsc{Uni})      & $\lambda_k = d_{max}(\epsilon + \frac{(k-1)(1-\epsilon)}{n-1})$  \\
		\hline
		\textsc{Geometric} (\textsc{Geo})    & $\lambda_k = d_{max}\epsilon^{\frac{n-k}{n-1}}$ \\
		\hline
		\textsc{(1-2-1)}  (\textsc{1-2-1})       & $\lambda_k = 2 - 2 \cos(\frac{\pi k}{n+1})$  \\
		\hline
		\textsc{Wilkinson} (\textsc{Wilk})    &   All positive, but one, roughly in pairs.\\
		\bottomrule
	\end{tabular}
\end{table}

The spectral properties of four types of artificial matrices
are given in Table \ref{spectral properties of generated matrices} and are explained below:

\begin{itemize}
    \item \textsc{Uniform} matrix: its
    eigenvalues are distributed equally within $[\min(d_{max}\epsilon,0), \max(d_{max}\epsilon, 0)]$ following a discrete uniform distribution.
    \item \textsc{Geometric} matrix: its spectrum follows a geometric distribution. If $d_{max}>0$ and $\epsilon \in (0,1)$, then its eigenvalues are in the range $(0, d_{max}\epsilon]$, and smaller eigenvalues are quite more clustered than the larger ones.
    \item \textsc{(1-2-1)} matrix \cite{gregory1978collection,higham1991algorithm} has analytically known eigenvalues. The clustering of its eigenvalues is not very strong, although clustering becomes tighter with the increase of dimension \cite{marques2008algorithm}.
    \item For \textsc{Wilkinson} matrix, all eigenvalues, but one, are positive. The positive eigenvalues are roughly in pairs, and the larger pairs are closer together.
\end{itemize}
Because of their distinct spectral proprieties, these 4 types of matrices should provide a qualitative picture of the behavior of the ChASE library resulting in widely different numerical responses and performance measurements.    
\begin{table*}[t]
	\renewcommand{\arraystretch}{1}
	\caption{Comparison of ChASE-CPU and ChASE-GPU with artificial matrices. The size of test matrices are $20\text{k}\times20\text{k}$, and {\sf nev} and {\sf nex} are $1500$ and $500$, respectively. Statistics for each test are obtained over {$20$} runs.}\label{Eigen type tests}
	\centering
\subfloat[ChASE-CPU on one node of JURECA-DC: MPI process number is $16$, and OpenMP thread number per rank is $8$.]{
\begin{tabular}{C{1.1cm}C{0.66cm}C{1.28cm}C{1.88cm}C{1.53cm}C{1.88cm}C{1.70cm}C{1.70cm}C{1.70cm}}
\toprule
  \multirow{2}{*}{Matrix} &
  \multirow{2}{*}{Iter.} &
  \multirow{2}{*}{Matvecs} &
  \multicolumn{6}{c}{Runtime (seconds)} \\ \cline{4-9} 
   &
   &
   &
  {\tt All} &
  \multicolumn{1}{c}{\tt Lanczos} &
  \multicolumn{1}{c}{\tt Filter} &
  \multicolumn{1}{c}{\tt QR} &
  \multicolumn{1}{c}{\tt RR} &
  \multicolumn{1}{c}{\tt Resid} \\ 
  \midrule

  \textsc{1-2-1}    & 13 & 466614 & $272.28 \pm 5.28$ & $4.64 \pm 0.19$ & $176.46 \pm 4.60$ & $31.69 \pm 1.27$ & $37.45 \pm 1.64$ & $20.99 \pm 0.67$ \\
  \textsc{Geo}      &  8 & 285192 & $165.39 \pm 1.86$ & $4.76 \pm 0.28$ & $108.02 \pm 1.75$ & $19.19 \pm 0.59$ & $20.64 \pm 1.22$ & $12.14 \pm 0.54$ \\
  \textsc{Uni}      &  5 & 163562 & $101.27 \pm 1.98$ & $4.76 \pm 0.24$ & $ 62.17 \pm 1.47$ & $12.05 \pm 0.53$ & $13.91 \pm 0.98$ & $ 7.97 \pm 0.60$ \\
  \textsc{Wilk}     &  9 & 248946 & $155.44 \pm 2.64$ & $4.86 \pm 0.96$ & $ 95.68 \pm 1.77$ & $21.53 \pm 0.88$ & $20.62 \pm 1.25$ & $12.09 \pm 0.47$ \\
  \bottomrule
\end{tabular}
}~\\

\subfloat[ChASE-GPU on one node of JURECA-DC: MPI process number is $4$, OpenMP thread and GPU number per process is $32$ and $1$.]{
\begin{tabular}{C{1.1cm}C{0.66cm}C{1.28cm}C{1.88cm}C{1.53cm}C{1.88cm}C{1.70cm}C{1.70cm}C{1.70cm}}
\toprule
  \multirow{2}{*}{Matrix} &
  \multirow{2}{*}{Iter.} &
  \multirow{2}{*}{Matvecs} &
  \multicolumn{6}{c}{Runtime (seconds)} \\ \cline{4-9} 
   &
   &
   &
  {\tt All} &
  \multicolumn{1}{c}{\tt Lanczos} &
  \multicolumn{1}{c}{\tt Filter} &
  \multicolumn{1}{c}{\tt QR} &
  \multicolumn{1}{c}{\tt RR} &
  \multicolumn{1}{c}{\tt Resid} \\ 
  \midrule
  
  \textsc{1-2-1}    & 13 & 466614 & $31.39 \pm 0.09$ & $0.58 \pm 0.01$ & $14.38 \pm 0.02$ & $2.59 \pm 0.01$ & $8.41 \pm 0.09$ & $5.24 \pm 0.04$ \\
  \textsc{Geo}      &  8 & 285192 & $18.57 \pm 0.05$ & $0.58 \pm 0.01$ & $ 8.76 \pm 0.02$ & $1.58 \pm 0.01$ & $4.58 \pm 0.04$ & $2.96 \pm 0.02$ \\
  \textsc{Uni}      &  5 & 163562 & $11.79 \pm 0.03$ & $0.58 \pm 0.01$ & $ 5.06 \pm 0.00$ & $1.00 \pm 0.00$ & $3.11 \pm 0.04$ & $1.96 \pm 0.02$ \\
  \textsc{Wilk}     &  8 & 246924 & $17.22 \pm 0.05$ & $0.57 \pm 0.00$ & $ 7.63 \pm 0.02$ & $1.59 \pm 0.00$ & $4.45 \pm 0.04$ & $2.90 \pm 0.02$ \\
  
  \bottomrule
\end{tabular}
}
\end{table*}

\begin{figure}[htbp]
\centering
\subfloat[Comparison of \texttt{Filter}'s performance in TFLOPS/node.]{\includegraphics[width=0.9\linewidth]{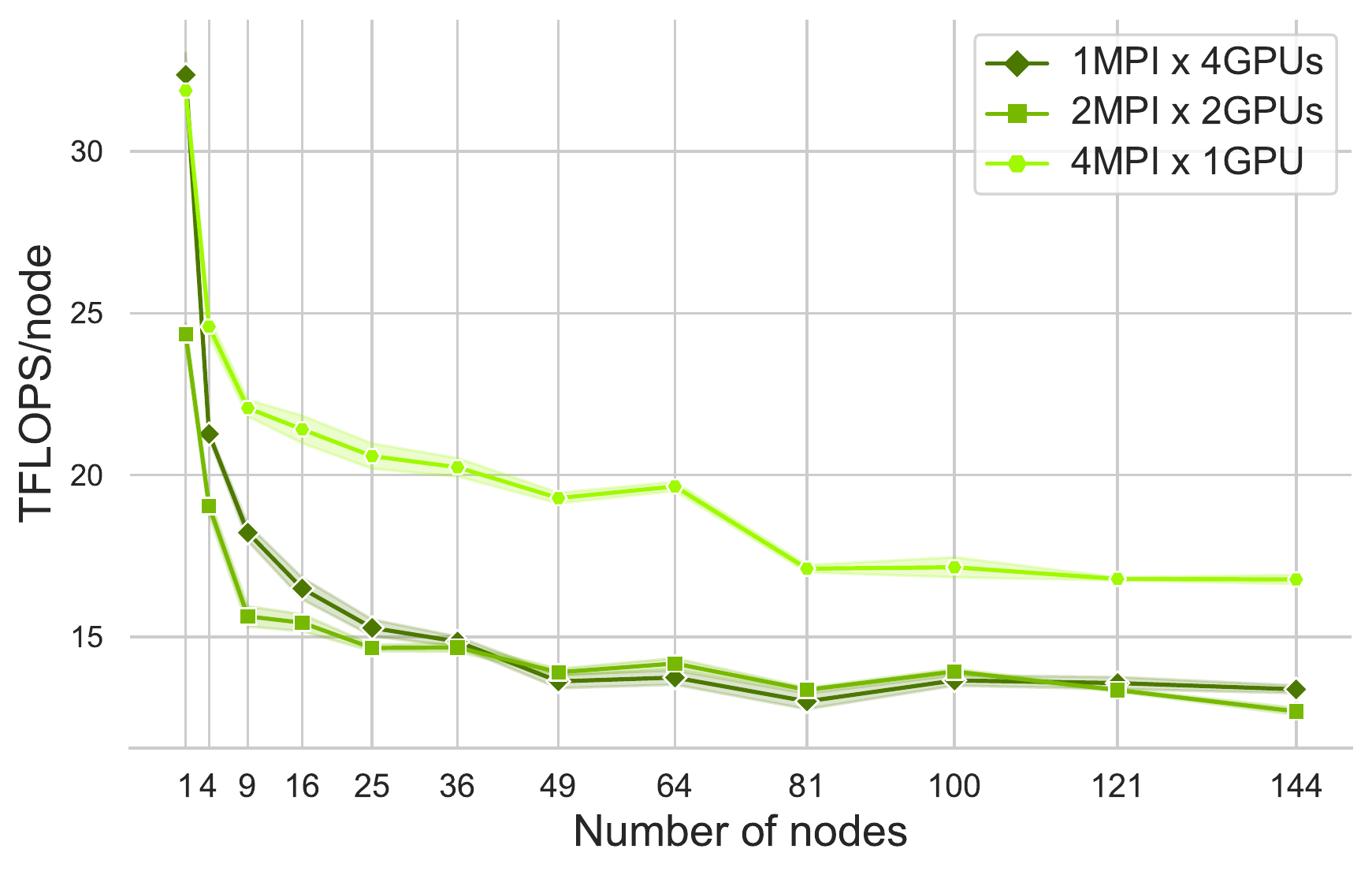}\label{fig:chase_weakscal_sweetspot_32threads_gpu_absolute}}
~\\
\subfloat[Comparison of time-to-solution of ChASE.]{\includegraphics[width=0.9\linewidth]{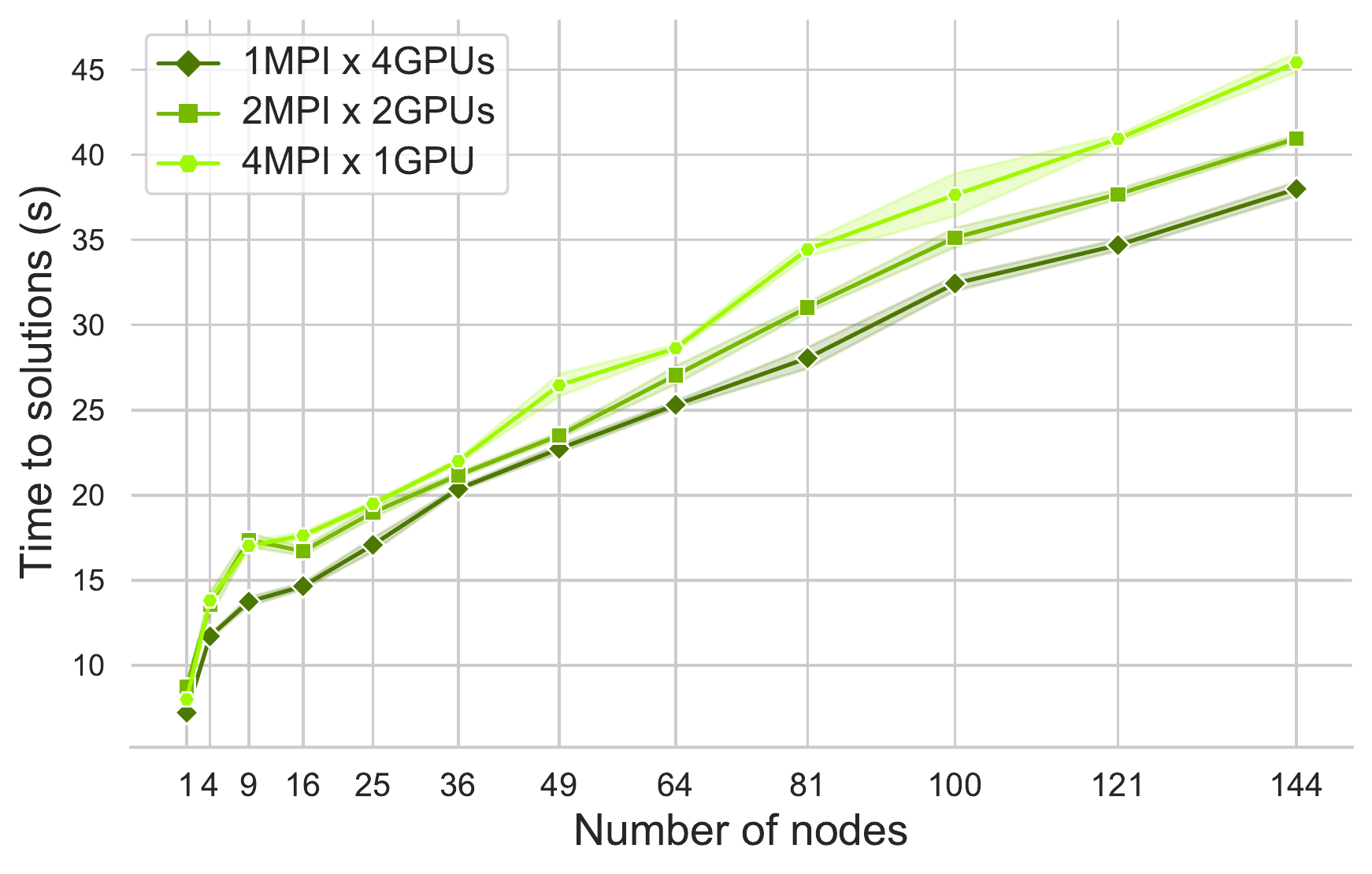}\label{fig:chase_weakscal_sweetspot_32threads_gpu_All}}
\caption{Evaluation of three MPI and GPU binding configurations: 1, 2 and 4 MPI rank with 32 threads each and 4 GPUs in total. Data are obtained as the averages of 20 repetitions.}
\end{figure}

\subsection{Evaluation of MPI and GPU Binding Configurations}\label{Evaluation of MPI and GPU Binding Configurations}

As shown in Section \ref{Multi-GPU Acceleration of ChASE}, ChASE-GPU supports 
a flexible binding policy of MPI ranks and GPUs. In order to find the best
configuration on the targeting platform JURECA-DC, we initially performed a weak scaling test
with three binding policies of MPI ranks and GPUs within node: {(1) 1 MPI rank bounded with 4 GPUs ($1$MPI$\times4$GPUs), (2) 2 MPI ranks with 2 GPUs bounded to each ($2$MPI$\times2$GPUs), (3) 4 MPI ranks with 1 GPU bounded to each ($4$MPI$\times1$GPUs)}.
The number
of threads per rank is fixed to $32$ so as to eliminate the huge NUMA-effects of some BLAS and
LAPACK routines.

For the weak scaling experiment, the numbers of compute nodes are
$p^2$, with $p \in 1, 2, \cdots, 12$ to produce 2D square node
grids. The generated test matrices are of type \textsc{Uniform} with
sizes being $3\times10^4p$. The number of desired eigenpairs {\sf nev} and external searching
space increment {\sf nex} are $2250$ and $750$, respectively. 

In this section, for all three configurations, we report both the performance of the {\tt Filter} and the time-to-solution of ChASE-GPU.
The {\tt Filter}, whose major part is the \textsc{HEMM}, is reported as the absolute performance extracted from the GPUs with tensor core activated. Because the FP64 Tensor Core is used automatically and selectively by cuBLAS 11.0.0 we report absolute performance and not fraction of peak.
The performance of the {\tt Filter} directly reflects the performance of the multi-GPU \textsc{HEMM}, which is one of the original contributions of this paper.
In the case of weak scaling, both the number of computing units and the problem size increase, which 
results in a constant workload per units. However, for an iterative eigensolver, it is impossible
to predict exactly the total workload to reach convergence, even if the problems are constructed
with matrices sharing the same spectral distribution.
Instead of solving problems to achieve full convergence, each test of
weak scaling has been executed with only one subspace iteration, which can ensure a constant
workload of Matvecs\footnote{It indicates the
  total number of matrix-vector multiplications executed by
  {\tt HEMM} within the {\tt Filter}.} per computing unit.

Fig. \ref{fig:chase_weakscal_sweetspot_32threads_gpu_absolute} shows that the performance of the {\tt Filter} decreases rapidly for all three configurations as the number of compute nodes increases, stabilising when the number of compute nodes is greater than 16. The reason for the performance drop is that communication (collective routine \textsc{MPI\_Allreduce}) and memory copies between CPU and GPU are included in the total execution time of the {\tt Filter}. In \cite{ZHANG2021108081} (see Supplementary Materials, Table S7), the authors showed that the latency in \textsc{MPI\_Allreduce} remains constant on more than 16 nodes, as does the impact of MPI communication on {\tt Filter} performance. This is clearly observed in the 1MPIx4GPU configuration when the number of nodes is increased from 1 to 4, as no MPI communication was required on one node (only 1 MPI rank is used).
At each step in the {\tt Filter}, a rectangular block of vectors $V$ is split (see Eq.~\ref{eq:distribution:AV}) and distributed to the GPUs of the same node and copied back to the host memory after the matrix multiplication is completed. These two operations (and $N=120k$) consume $~30\%$ of the total time of the distributed {\tt HEMM}. In addition, some extra time ($~19\%$) is spent on inter-GPU communication at the node level, so up to $50\%$ of the {\tt HEMM} time is spent on the memory copy. The memory copies cannot be efficiently overlapped with matrix-matrix multiplication because there is a strong dependency between them. 
Currently, this multi-GPU {\tt HEMM} lacks support for faster communication links between GPUs within the node, such as. \textsc{NVLink}, and is part of future work.

{Fig.~\ref{fig:chase_weakscal_sweetspot_32threads_gpu_All}} shows that the time-to-solution for ChASE with all three configurations
increases somewhat linearly as a function of the number of compute nodes used. 
The performance of the entire ChASE is different from the performance of the {\tt Filter} in Fig. \ref{fig:chase_weakscal_sweetspot_32threads_gpu_absolute}.
ChASE with configuration $1$MPI$\times4$GPUs always outperforms the other two, with $2$MPI$\times2$GPUs in between. 
Since QR and RR are computed redundantly on each MPI rank and operate on the full column size, the gain of the configuration with $1$MPI$\times4$GPUs over the other configurations comes from a lower communication overhead using expensive \textsc{MPI\_Ibcast} (see \cite{ZHANG2021108081}, Supplement Materials, Table S7). Unlike {\sc MPI\_Allreduce}, the latency of the broadcasting routines increases steadily with the number of MPI ranks.


The outcome of its higher efficiency due to the decreased MPI communication makes up the relative lower performance of its corresponding {\tt Filter}.
Because $1$MPI$\times4$GPUs is the best configuration of ChASE-GPU on JURECA-DC, we will use it 
as the default configuration for the remaining tests. Additional weak
scaling tests are discussed in Section \ref{Weak scaling Performance}.

\subsection{Eigen-type tests}

In order to confirm the numerical robustness of ChASE-GPU, we compare it with
ChASE-CPU using the test matrix suite described in  
Section \ref{Random Matrices with Given Spectra}. The size of the test matrices is
fixed as $20$k, and {\sf nev} and {\sf nex}
are respectively $1500$ and $500$, which means the maximum size of active
subspace is $10$\% of the full space of problems. The $\ell^2$-norm condition numbers of generated \textsc{(1-2-1)}, \textsc{Geometric}, \textsc{Uniform} and \textsc{Wilkinson} matrices are respectively $1.6\times 10^8$, $1.0\times 10^4$, $1.0\times 10^4$ and $4.7\times 10^4$.

The experiments of both ChASE-CPU and ChASE-GPU are performed
on one single compute node with a different combination of MPI ranks and OpenMP
threads for each of the two versions of ChASE. For ChASE-CPU,
the number of MPI ranks and OpenMP threads per node is fixed at $16$
and $8$, respectively. This is the best combination of MPI and OpenMP on JURECA-DC, and
was obtained from a series of sweet-spot tests spanning all possible combinations.
For ChASE-GPU, the configuration is {$1$MPI$\times4$GPUs}, and the number of OpenMP threads per MPI rank is
$32$, which has been proved as the best one. 

The results are shown in Table \ref{Eigen type tests}, which includes
the subspace iteration number until convergence, 
the required number of Matvecs operations, and the runtime for ChASE
and its main parts. For all four types of eigenproblems, both ChASE-CPU
and ChASE-GPU are able to achieve the convergence in a limited number
of iterations with the \textsc{(1-2-1)} problem, {which has a much larger condition number}, taking the most time {and iterations}, more than doubling the runtime and {iterations} of the \textsc{Uniform} problem. The acceleration provided by ChASE-GPU is practically independent from the type of eigenproblem. For all four test matrices, ChASE-GPU achieves a speedup of approximately {$8.9\times$} for the entire runtime and {$12.7\times$} for just the {\tt Filter}, which is the most computationally intensive part of the solver. The considerations above demonstrate the viability of ChASE as a general purpose solver for extremal symmetric eigenproblems. Because they converge faster than the others, we will generate only eigenproblems of the \textsc{Uniform} type for the scalability tests. 

A closer look reveals that the exact number of iterations between ChASE-CPU and ChASE-GPU
differs for the matrix \textsc{Wilkinson}. 
This difference is also reflected in the numbers of Matvecs. This may
seem an harmless difference, but it is rather suspicious in light of the
deterministic convergence provided by the Chebyshev filter~\cite{Winkelmann2019}.  
Upon further investigation, we identified the cause of this behavior in
a very peculiar numerical instability of {\tt cusolverXgeqrf}, the QR factorization
of cuSOLVER, which seems to happen randomly. 
The numerical difference of QR factorization between the one in cuSOLVER and LAPACK
is minor, just above the machine precision. However, this difference propagates through
the computation, which finally results in a slightly different numerical
accuracy. We further observed that for much larger matrices than $20$k
such numerical instability can sometimes damage the redundant computations of the
QR factorization and introduce a mismatch in the data exchanged between
different rows of MPI communicators. Eventually this behavior results
in the breakdown of ChASE-GPU. We have signalled the bug to the NVIDIA
developers of cuSOLVER.  


\subsection{Scalability}

This section analyze ChASE-GPU's behavior in strong and weak scalability regime by
comparing with ChASE-CPU. For all the tests, the numbers of MPI ranks and OpenMP threads per rank of ChASE-CPU are respectively $16$ and $8$.
For ChASE-GPU, the number of MPI ranks per node is {$1$, with $4$ GPUs and $32$}
threads assigned to each rank. 

\begin{figure}[t]
\centering
\subfloat[Strong scaling performance of ChASE-CPU.]{\includegraphics[width=0.9\linewidth]{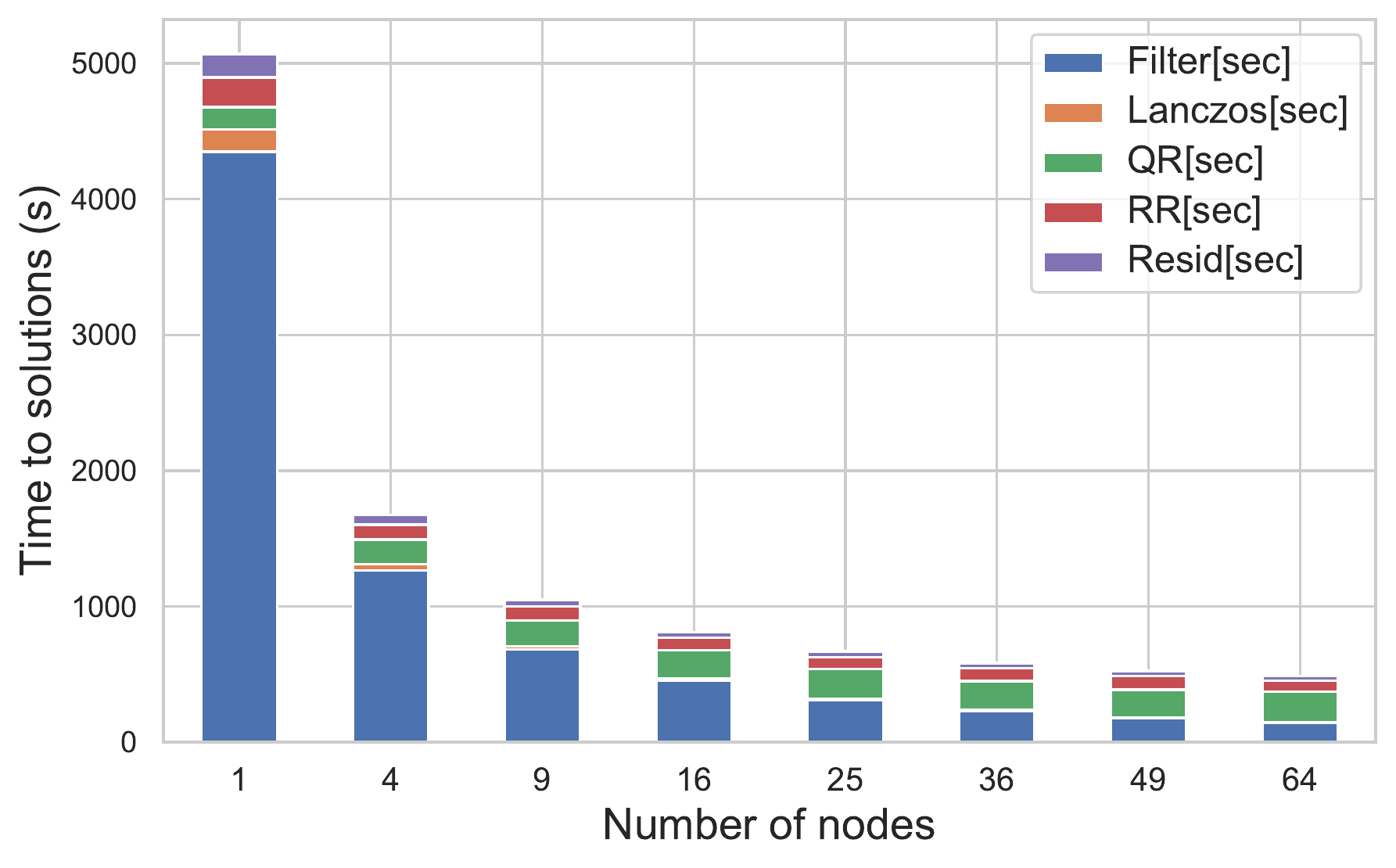}\label{fig:strongscaling_timings_cpu}}
~\\
\subfloat[Strong scaling performance of ChASE-GPU.]{\includegraphics[width=0.9\linewidth]{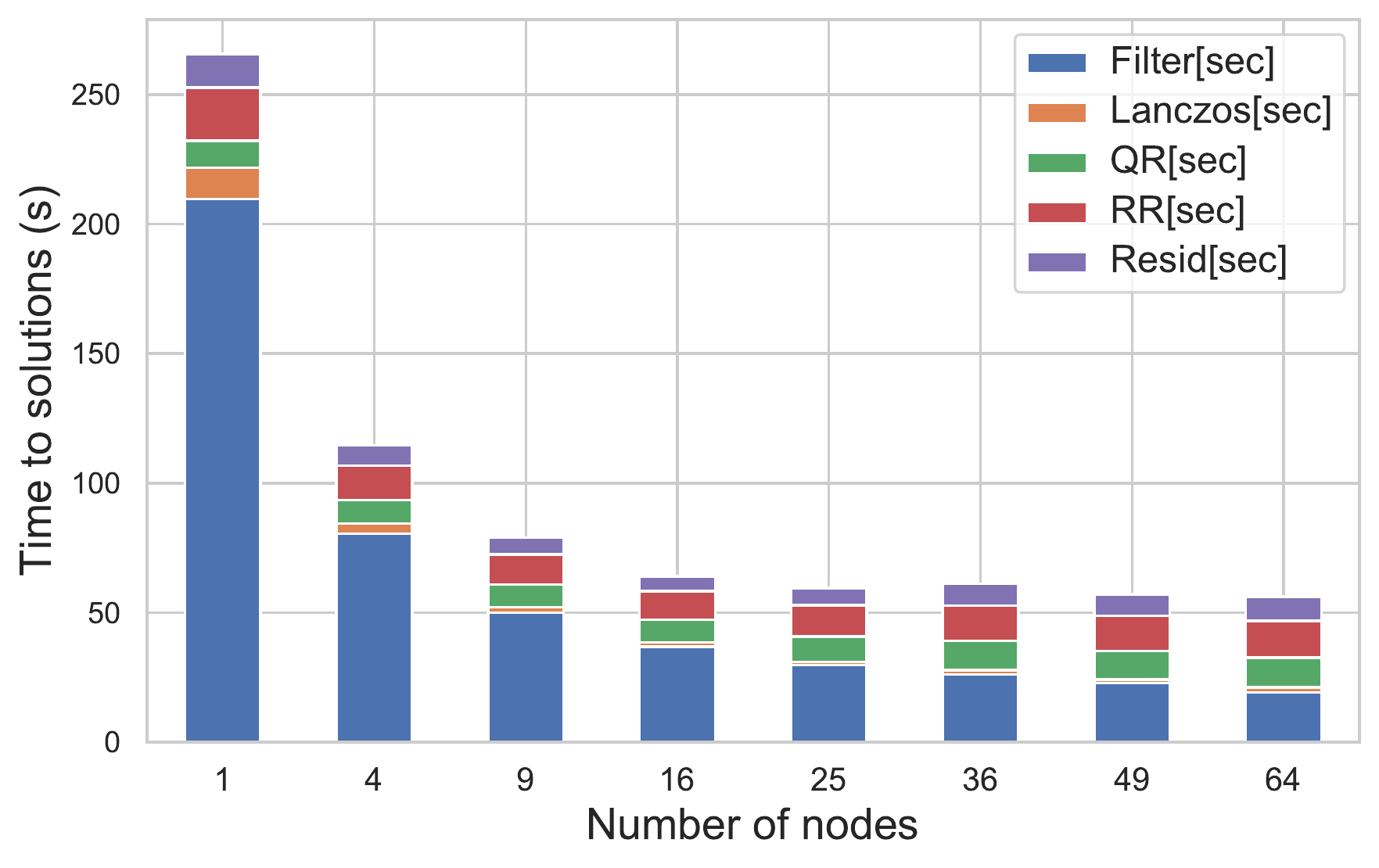}\label{fig:strongscaling_timings_gpu}}
\caption{Strong scaling tests with \textsc{Uniform} matrix ($n=130,000$, {\sf nev}$=1000$, and {\sf nex}$=300$). Data are obtained as the averages of 15 repetitions.}\label{fig:strong scaling timings}
\end{figure}

\subsubsection{Strong scaling}

\begin{figure}[t]
\centering
\includegraphics[width=0.9\linewidth]{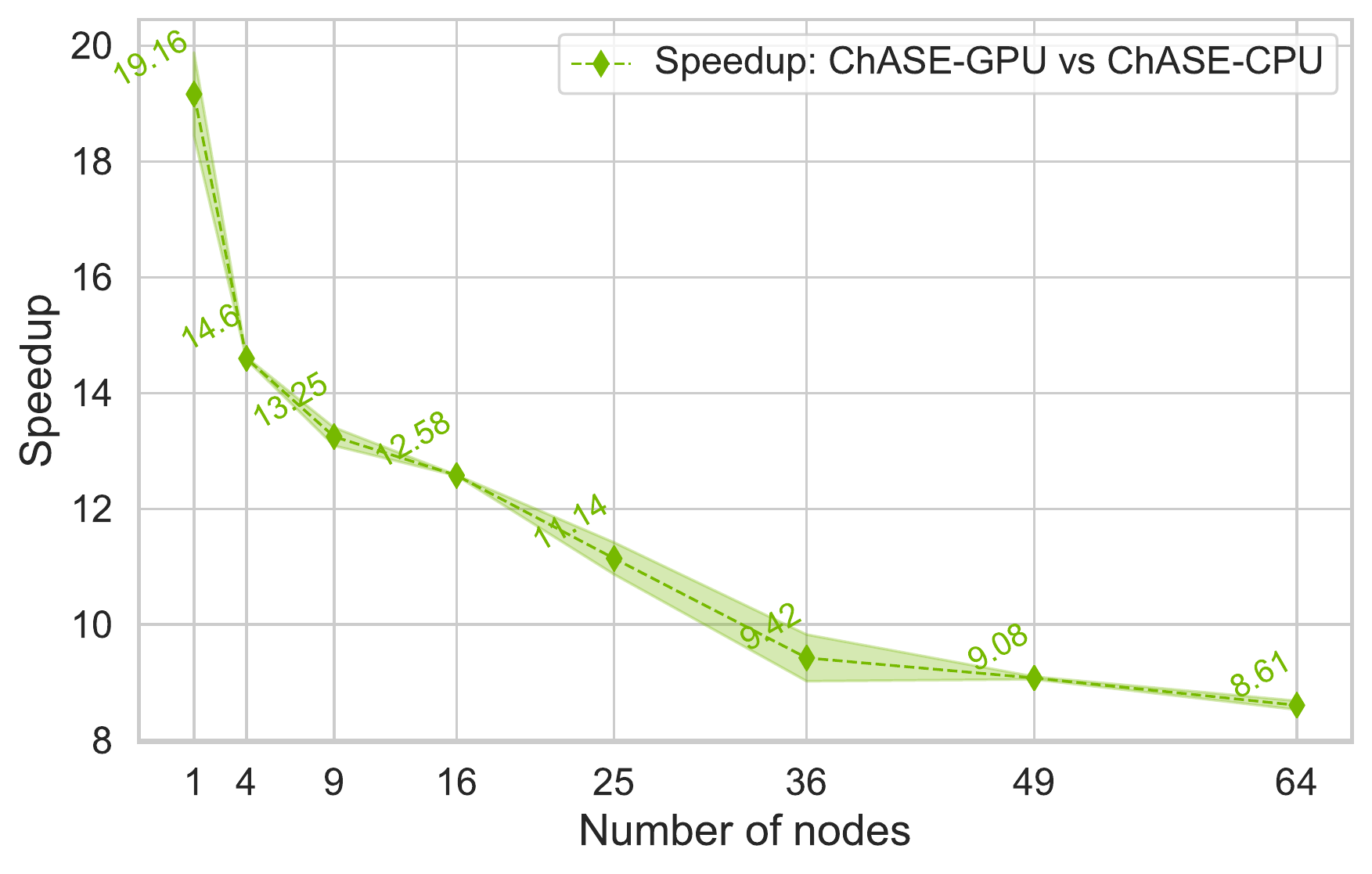}
\caption{Strong scaling: Speedup. Speedup of ChASE-GPU over ChASE-CPU. Error bars are obtained with 15 repetitions.}\label{fig:strongscaling_speedup}
\end{figure}

Fig. \ref{fig:strong scaling timings} illustrates the results of the strong scaling experiment 
of ChASE-CPU and GPU using a \textsc{Uniform} matrix of size $n=130,000$. We fix
{\sf nev} and {\sf nex} respectively as $1000$
and $300$ ($=1$\%n).
The counts of compute nodes are selected to be square numbers $1, 4, 9, \cdots, 64$.
Fig. \ref{fig:strong scaling timings} reports the runtime
of ChASE-CPU and ChASE-GPU as a vertical stacked bar plot, which includes also the fractions of
runtime of numerical functions, such as {\tt Filter}, \texttt{Lanczos}, \texttt{QR}, \texttt{RR} and \texttt{Resid}. The speedup of ChASE-GPU is plotted in Fig. \ref{fig:strongscaling_speedup}, where for each point on the x-axis, the speedup is calculated with respect to the corresponding timing of ChASE-CPU.

Both ChASE-CPU and ChASE-GPU
can achieve good strong scaling performance for smaller number of nodes. However, with larger number of compute nodes, the decrease of total runtime of ChASE become progressively negligible, especially for ChASE-GPU.
The \texttt{Filter}, whose most important operation is the customized
\texttt{HEMM}, achieves very good strong scaling performance in both ChASE-CPU and ChASE-GPU. Compared with
the tests using $1$ compute node, {ChASE-CPU with $64$ compute nodes achieves $32\times$ 
speedup for \texttt{Filter}, $29\times$ speedup for \texttt{Lanczos}, and $5\times$ speedup 
for \texttt{Resid}. Analogously, ChASE-GPU achieves $10.8\times$ speedup for \texttt{Filter}, $5\times$ speedup for \texttt{Lanczos},  but only $1.4\times$ speedup for \texttt{Resid}.} For ChASE-CPU,
the most dominant linear algebra operation in the \texttt{Filter}, \texttt{Lanczos} and \texttt{Resid} is \texttt{HEMM}.
In these three functions in ChASE-GPU, only \texttt{HEMM} has been offloaded
to GPUs, which achieves a notable acceleration over the CPU version. Compared to \texttt{HEMM}, the remaining BLAS/LAPACK operations called within the \texttt{Lanczos} and \texttt{Resid} become much more dominant, which turn them into new bottlenecks. 
This is also the reason why the strong scaling performance
of ChASE-GPU tends to be worse than ChASE-CPU, even with the acceleration of GPUs.
This is clearly visible from Fig.~\ref{fig:strongscaling_speedup}, which shows the speedup of ChASE-GPU over ChASE-CPU as a function of compute nodes count.
ChASE-GPU with $1$ compute node has the maximal speedup over ChASE-CPU, which is {$19.16$}. Increasing the count of compute node, the speedup keeps getting smaller and tends to flatten towards a value {$\sim8.61$}.

\subsubsection{Weak scaling}\label{Weak scaling Performance}

\begin{figure}[t]
\centering
\subfloat[Weak scaling performance of ChASE-CPU.]{\includegraphics[width=0.9\linewidth]{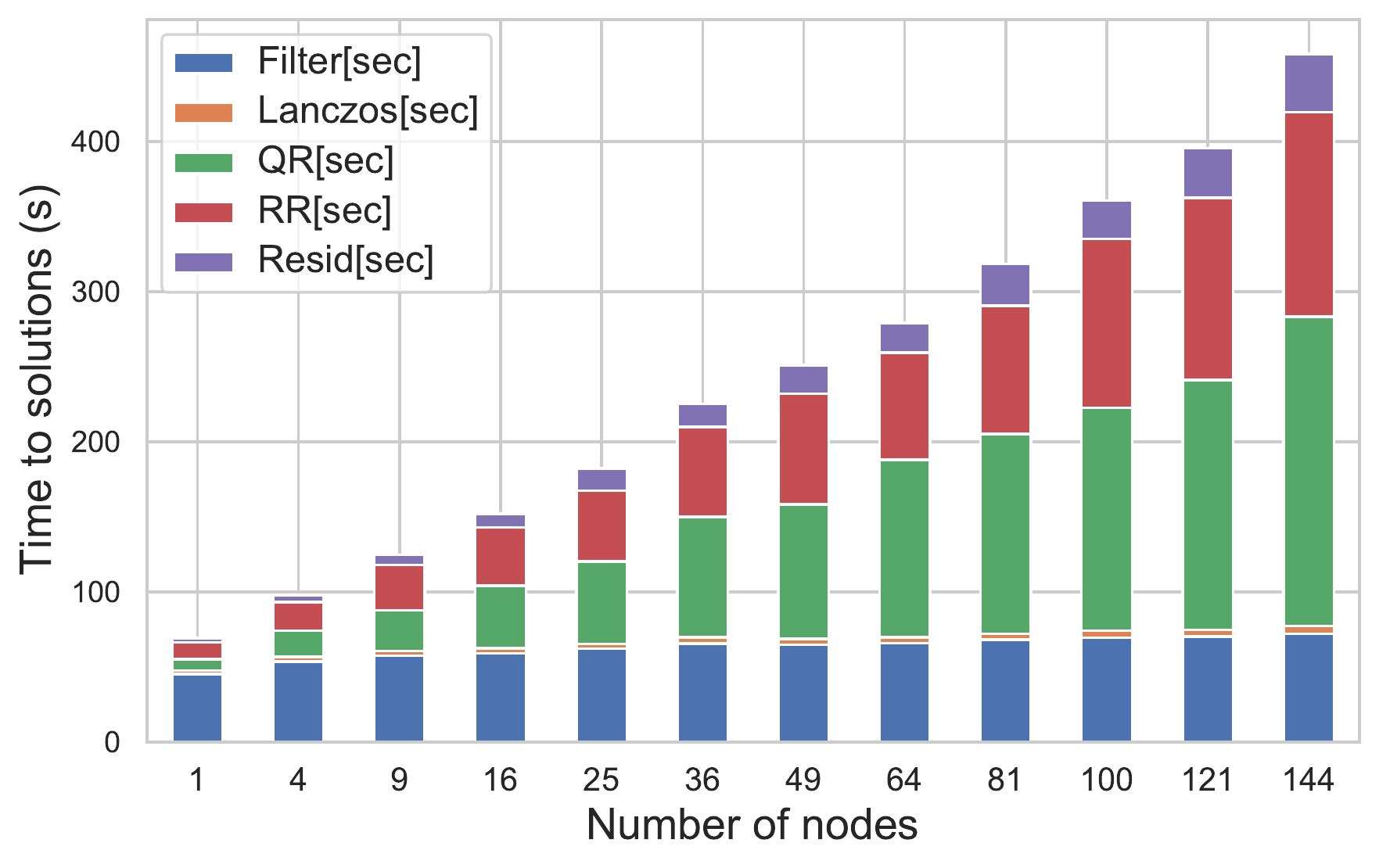}\label{fig:weakscaling_timings_cpu}}
~\\
\subfloat[Weak scaling performance of ChASE-GPU.]{\includegraphics[width=0.9\linewidth]{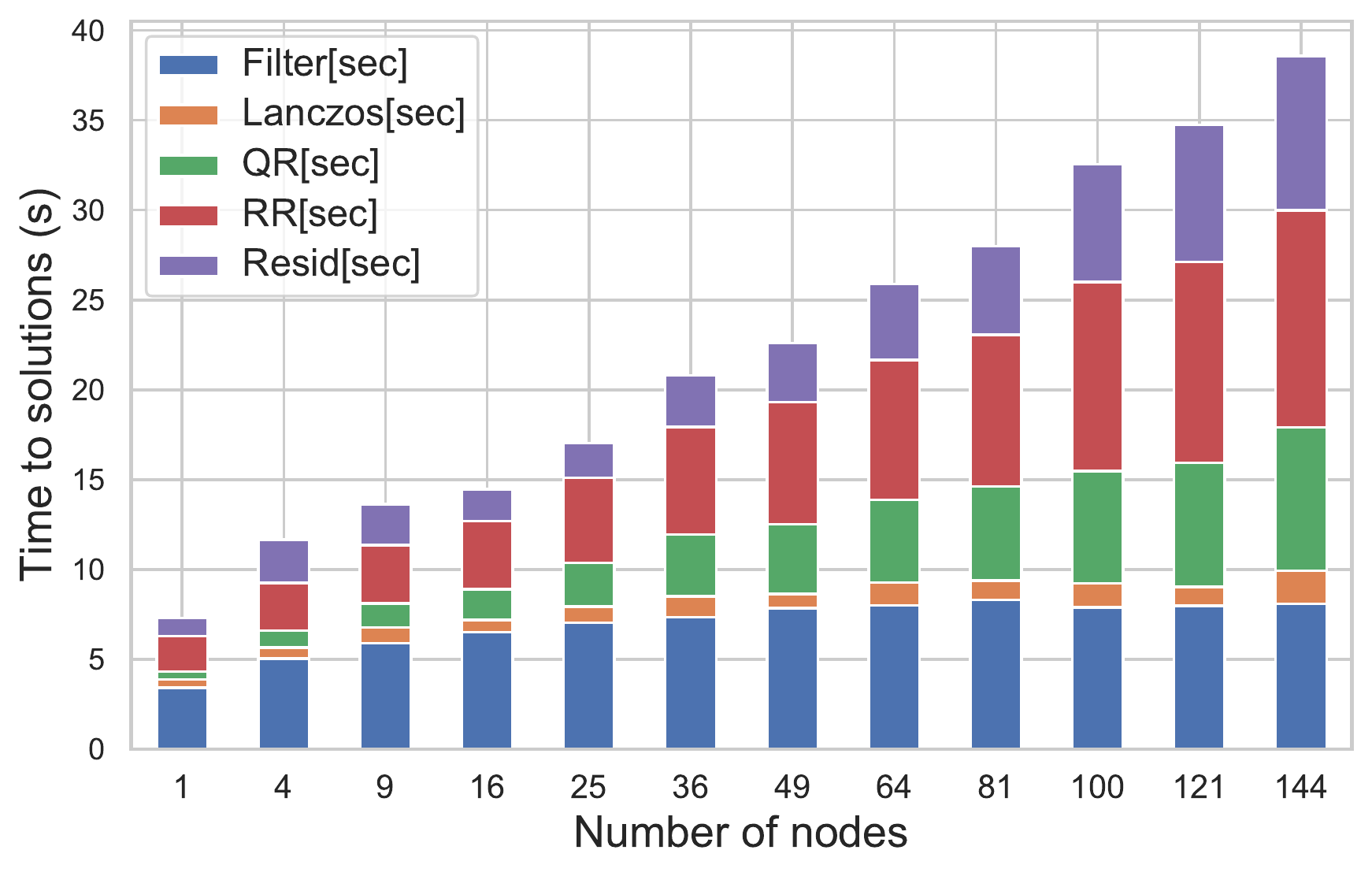}\label{fig:weakscaling_timings_gpu}}
\caption{Weak scaling tests with \textsc{Uniform} matrix ($n$ ranging from $30$k to $360$k, {\sf nev}$=2250$,  {\sf nex}$=750$). Data are obtained as the averages of 15 repetitions. }\label{fig:weakscaling_timings}
\end{figure}

\begin{figure}[t]
\centering
\includegraphics[width=0.9\linewidth]{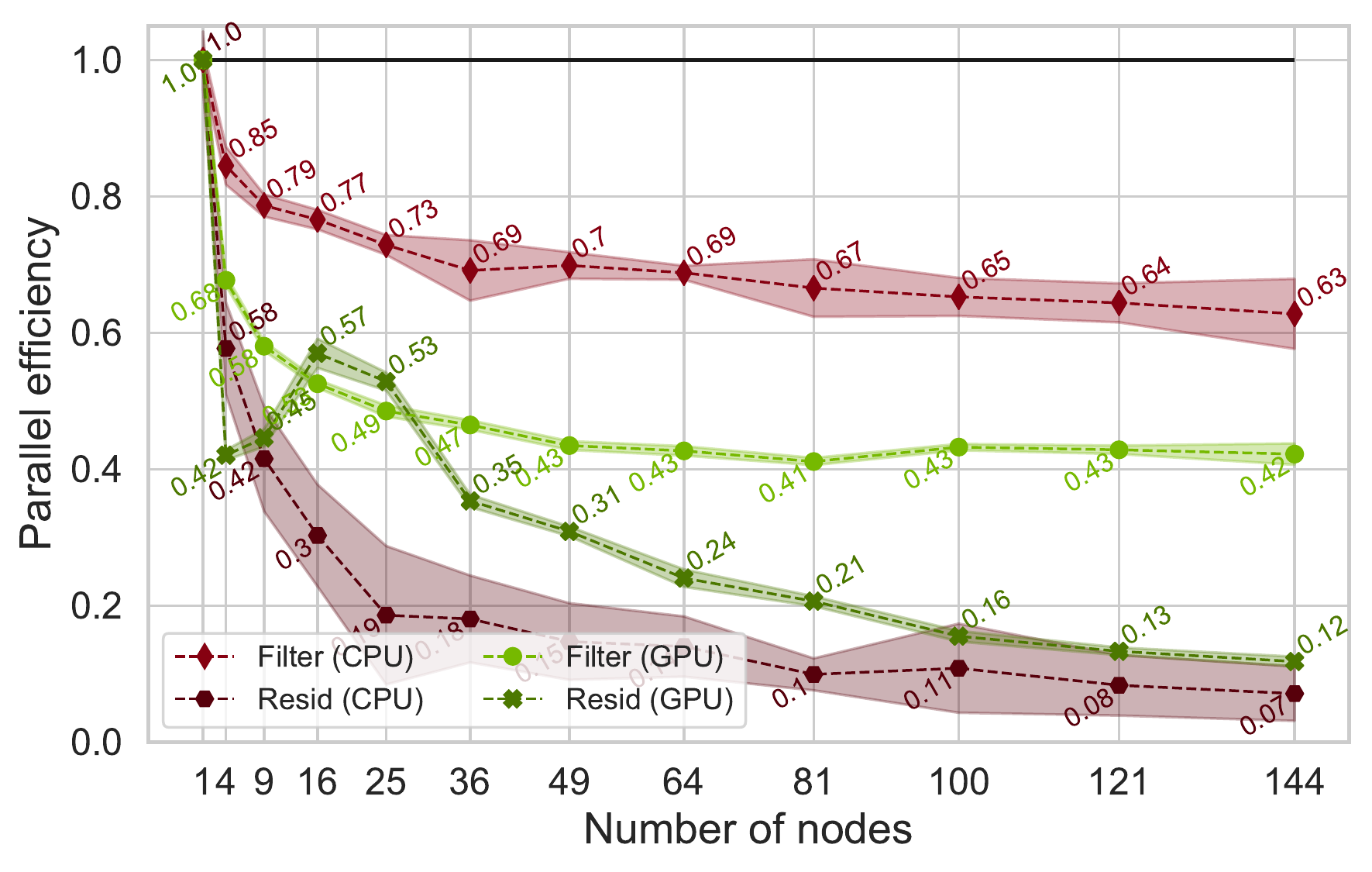}
\caption{Weak scaling: Parallel efficiency of \texttt{Filter} and \texttt{Resid}. Error bars are obtained with 15 repetitions.}\label{fig:weakscaling_eff_Filter_Resid}
\end{figure}

Weak scaling experiments are particularly important to domain scientists, who are interested in simulating system of increasingly larger size. In order to maintain a fixed workload, we keep the same setup described 
in Section \ref{Evaluation of MPI and GPU Binding Configurations}.
The test matrices are of type \textsc{Uniform}, with size increment of $30$k ($30\text{k}, 60\text{k}, 90\text{k}, \cdots, 360\text{k}$). 
The counts of compute nodes selected as square
numbers $1, 4, 9, \cdots, 144$, and {\sf nev} and {\sf nex} are respectively fixed as $2250$ and $750$.
Fig. \ref{fig:weakscaling_timings} plots the results of weak scaling experiments
as a vertical stacked bar plot, which shows the runtime of ChASE-CPU and ChASE-GPU, including
the runtime of their numerical functions. Additionally, Fig. \ref{fig:weakscaling_eff_Filter_Resid}
reports the parallel efficiency of the numerical functions \texttt{Filter} and \texttt{Resid}
of this weak scaling experiment.

The good news is that, independently of which version, ChASE scale linearly. The bad news is that the total runtime of ChASE-CPU and ChASE-GPU doubles every-time the matrix size {quadruples and triples, respectively}. 
When we look at the details of the distribution of runtime over the different functions, we observe 
a good weak scaling of the \texttt{Filter} thanks to the custom parallelization of \texttt{HEMM} for both CPU and GPU. 
However, a small increase in {\tt Filter} runtime is observed when the number of nodes is increased, e.g. to 1, 4 and 16 nodes (Fig. 5b) the runtime is ~3.27 sec, ~5.01 sec and ~6.3 sec, respectively. The main reason for this is an increased amount of communication ({\sc MPI\_Allreduce}). The percentage of MPI on 4 and 16 nodes is 35\% and 49\% of the total Filter execution time, respectively. However, considering only the distributed {\tt HEMM} performance (without MPI communication) on 16 nodes with 64 GPUs, we reach 685.44 TFlops (~55\% of the peak GPU performance). The increased communication is expected because Allreduce is called at the end of the distributed {\tt HEMM} which is computed multiple times within the Filter and could be repeated up to $20$ times (the maximum degree of the polynomial) in the first iteration of ChASE.

The weak scaling of \texttt{Lanczos} and \texttt{Resid} are quite worse
than the \texttt{Filter}, even if they make use of the distributed \texttt{HEMM}. 
With the increase
of problem size, \texttt{QR} and \texttt{RR}, which are computed redundantly on the node, become progressively dominant, especially the QR factorization. For ChASE-CPU with $144$ compute nodes, \texttt{QR}
and \texttt{RR} take {$48$\% and $29$\%} of the whole runtime, meanwhile \texttt{Filter} take only
{$16$\%}. In ChASE-GPU, the QR factorization and the \texttt{GEMM} routine, called internally by \texttt{RR}, have been offloaded
to a single GPU using the cuSOLVER and the cuBLAS libraries. Such a choice makes these two functions less impactful than the corresponding one in ChASE-CPU.

Fig. \ref{fig:weakscaling_eff_Filter_Resid} shows the parallel efficiency of \texttt{Filter} and \texttt{Resid} of both ChASE-CPU and ChASE-GPU. For $144$ compute nodes, the \texttt{Filter}
in ChASE-CPU and ChASE-GPU shows a parallel efficiency of {$63$\% and $42\%$}, respectively. On the other hand, the parallel
efficiency of \texttt{Resid} in ChASE-CPU and ChASE-GPU attains {$7$\% and $12$\%},
respectively.
Overall, our results confirms the efficiency of our implementation of distributed multi-GPU \texttt{HEMM} and provide a strong indication of what should be the focus of further developments of the ChASE library. 
 

\begin{figure}[t]
\centering
\includegraphics[width=0.9\linewidth]{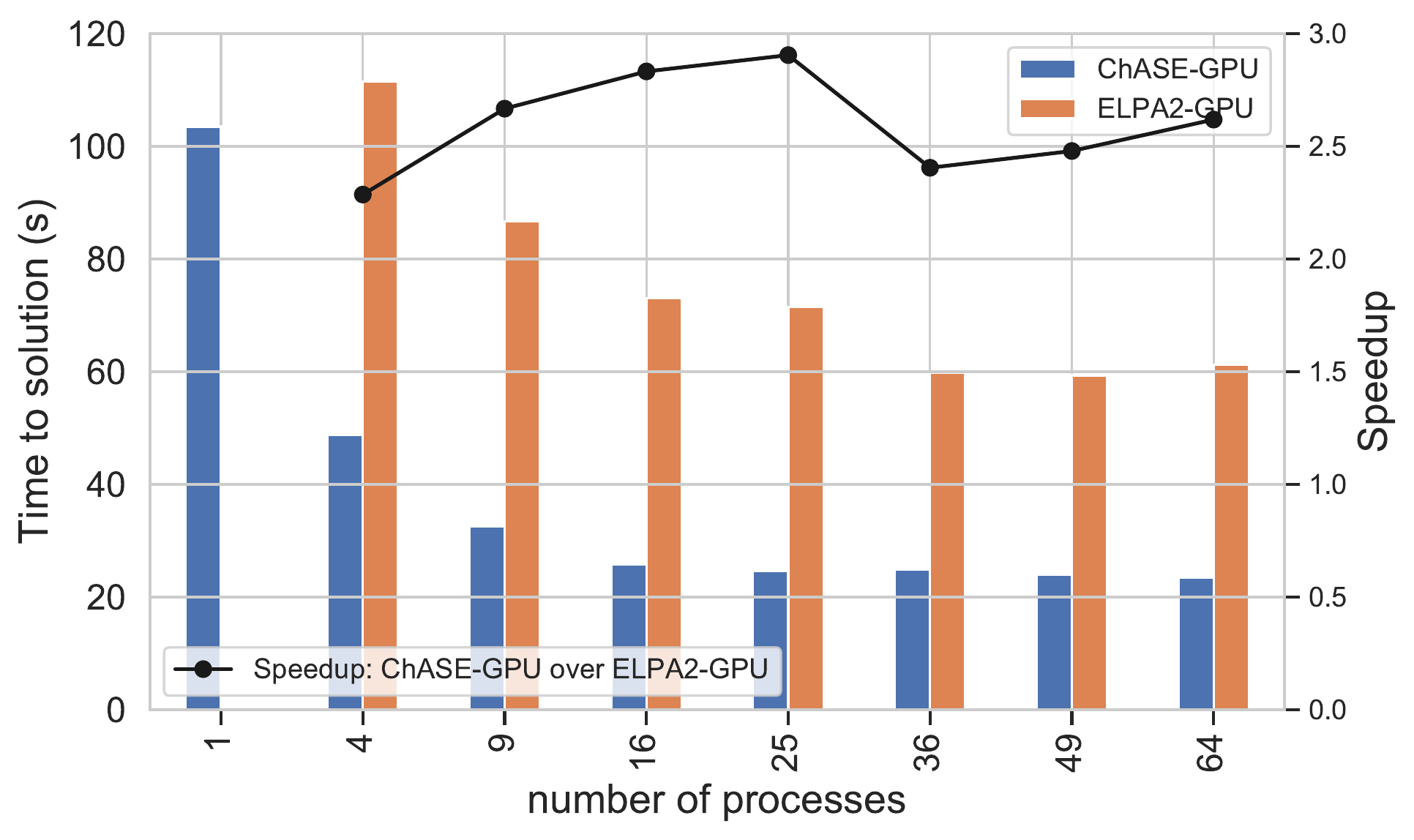}
\caption{Strong scaling: Time-to-solution and speedup of ChASE-GPU over ELPA2 for solving 76k In$_2$O$_3$ Hermitian eigenproblem with {\sf nev=800}. Data are obtained as the averages of 15 repetitions.}\label{fig:chase_vs_elpa_gpu}
\end{figure}

\subsection{Comparison with other libraries}

As we stated in Section \ref{Distributed eigensolvers}, there are no other distributed GPU eigensolvers apart from ELPA2. Therefore, we perform a 
strong scaling test up to 64 compute nodes comparing ChASE-GPU with ELPA2 with GPU support (ELPA2-GPU). 
The comparison of ChASE-CPU with other libraries are not carried out in this paper, since the comparison with ScaLAPACK, Elemental and FEAST are available in \cite{Winkelmann2019}.
The version of ELPA for the benchmarks is 2020.11.001, which is the most updated installation on JURECA-DC. The selected installation is 
compiled with GCC 10.3.0, OpenMPI 4.1.1, Intel MKL 2021.2.0 and CUDA 11.3 with CUDA architecture $sm\_80$. We prefer to use ELPA compiled with OpenMPI rather
than the one compiled with ParaStationMPI, since the former enable ELPA to be 10\% faster than the latter. The Multi-Process Service (MPS) is activated for ELPA. The MPI core and GPU numbers per node is 
set respectively as $32$ and $4$. This configuration has been selected based on a sweet-spot test with multiple configurations. 
The 2D grid of MPI ranks is setup as closest to be square. The block size of the block-cyclic distribution of matrix in ELPA is fixed at $16$.

The eigenproblem that we use for this test is Hermitian with a matrix size 76k, and is generated by the discretization of the Bethe-Salpeter equation used to simulate the opto-electornic properties of In$_2$O$_3$. The number
of eigenpairs sought after, {\sf nev}, is set at $800$ for both ChASE-GPU and ELPA2-GPU. For ChASE-GPU, the size of the external searching space {\sf nex} is fixed as $200$. 
The time-to-solution and speedup of ChASE-GPU over ELPA2-GPU is reported in Fig. \ref{fig:chase_vs_elpa_gpu}.

We first point out that ELPA2-GPU runs out of device memory when only 1 compute node is used, while ChASE-GPU solves successfully the problem in $104$ seconds. The strong scaling performance of ChASE-GPU is also better than the one of ELPA2-GPU, especially with a relative small number of compute nodes. For instance, ChASE-GPU shows a $1.88\times$ speedup when the compute node number increases from $4$ to $16$, meanwhile ELPA2-GPU displays only $1.54\times$ speedup. In average, ChASE-GPU achieves $2.6\times$ speedup over ELPA2 when the compute node number ranges from $4$ to $16$. The maximal speedup
$2.97\times$ has been achieved when $25$ compute nodes are used.

We point out that the performance gain of ChASE-GPU over ELPA2-GPU has been obtained when only a relatively small portion of extremal eigenpairs are sought after, which is the range of viability of the ChASE library. In this case, ChASE-GPU can achieve large speedup over ELPA2-GPU with an inferior memory footprint.

\section{Conclusion}\label{Conclusion}

In this paper, we presented a distributed CPU-GPU implementation of the ChASE
eigensolver for large-scale symmetric eigenproblems. ChASE targets extremal dense eigenproblems
when a relatively small fraction ($\leq10$\%) of extremal eigenpairs is sought after. We introduce
the implementation of a customized distributed CPU-GPU \texttt{HEMM} for ChASE which is used in many of its functions, notably the Chebyshev filter.
Because the {\tt Filter} function is the most computationally heavy part of the library, this custom-HEMM implementation has a dramatic impact on the parallel performance of the library when is ported on distributed multi-GPUs architectures.
We have benchmarked the numerical and parallel performance of the new distributed hybrid CPU-GPU implementation on one of the most modern platforms featuring AMD Epyc Rome CPUs coupled with 4 powerful NVIDIA A100. Our tests show a good parallel performance of the
custom \texttt{HEMM} implementation impacting positively the overall performance of the library. Because of the excellent scaling of the {\tt Filter} using the new HEMM, other functions in the library have become the new bottleneck 
which we plan to addressed in the near future. 
The overall target, is to further develop ChASE into an eigensolver that can be deployed and used on current PETAscale supercomputing clusters to solve for very large eigenproblems.

\section{Acknowledgments}

This work was supported by the Croatian Science Foundation under grant number HRZZ-UIP-2020-02-4559, by the Ministry of Science and Education of the Republic of Croatia and the Deutsche Akademische Austauschdienst (DAAD) from fund of the Bundesministeriums für Bildung und Forschung (BMBF) through project "PPP Kroatien" ID 57449075. This work was also partially supported by PRACE-6IP WP8: Performance Portable Linear Algebra (grant agreement ID 823767).  

\bibliographystyle{ACM-Reference-Format}
\bibliography{references.bib}

\end{document}